\documentclass[aps,prb,twocolumn,superscriptaddress]{revtex4-1}

\usepackage[pdftex]{graphicx}
\usepackage{color,soul}
\usepackage{enumitem}
\usepackage{gensymb}
\usepackage{epstopdf}
\usepackage{bm}
\usepackage{amsmath,amsthm,amssymb}
\usepackage{dcolumn}

\begin{document}


\title{Reorientations, relaxations, metastabilities, and multidomains of skyrmion lattices}



\author{L.J. Bannenberg}
\affiliation{Faculty of Applied Sciences, Delft University of Technology, Mekelweg 15, 2629 JB Delft, The Netherlands}
\email{l.j.bannenberg@tudelft.nl}
\author{F. Qian}
\affiliation{Faculty of Applied Sciences, Delft University of Technology, Mekelweg 15, 2629 JB Delft, The Netherlands}
\author{R.M. Dalgliesh}
\affiliation{ISIS neutron source, Rutherford Appleton Laboratory, STFC, OX11 0QX Didcot, United Kingdom}
\author{N. Martin}
\affiliation{Laboratoire L\'{e}on Brillouin, CEA-Saclay, 91191 Gif sur Yvette, France}
\author{G. Chaboussant}
\affiliation{Laboratoire L\'{e}on Brillouin, CEA-Saclay, 91191 Gif sur Yvette, France}
\author{M. Schmidt}
\affiliation{Max Planck Institute for Chemical Physics of Solids, 01187 Dresden, Germany}
\author{D.L. Schlagel}
\affiliation{Ames Laboratory, Iowa State University, Ames, IA 50011, USA}
\author{T.A. Lograsso}
\affiliation{Ames Laboratory, Iowa State University, Ames, IA 50011, USA}
\affiliation{Department of Materials Science and Engineering, Iowa State University, Ames IA 50011 USA}
\author{H. Wilhelm}
\affiliation{Diamond Light Source Ltd., OX11 0DE Didcot, United Kingdom}
\author{C. Pappas}
\affiliation{Faculty of Applied Sciences, Delft University of Technology, Mekelweg 15, 2629 JB Delft, The Netherlands}

\date{\today}

\begin{abstract}
Magnetic skyrmions are nano-sized topologically protected spin textures with particle-like properties. They can form lattices perpendicular to the magnetic field and the orientation of these skyrmion lattices with respect to the crystallographic lattice is governed by spin-orbit coupling. By performing small angle neutron scattering measurements, we investigate the coupling between the crystallographic and skyrmion lattices in both Cu$_2$OSeO$_3$ and the archetype chiral magnet MnSi. The results reveal that the orientation of the skyrmion lattice is primarily determined by the magnetic field direction with respect to the crystallographic lattice. In addition, it is also influenced by the magnetic history of the sample which can induce metastable lattices. Kinetic measurements show that these metastable skyrmion lattices may or may not relax to their equilibrium positions under macroscopic relaxation times. Furthermore, multidomain lattices may form when two or more equivalent crystallographic directions are favored by spin-orbit coupling and oriented perpendicular to the magnetic field.  
\end{abstract}

\maketitle

\section{Introduction}
Magnetic skyrmions are nano-sized topologically protected spin textures with particle-like properties which may form lattices oriented perpendicular to the magnetic field.\citep{muhlbauer2009,fert2013,nagaosa2013,milde2013,romming2013,seki2015skyrmions,bauer2016generic} These Skyrmion Lattices (SkL) were first identified in cubic chiral magnets by Small Angle Neutron Scattering (SANS) inside the A-phase, which is a pocket in the magnetic field ($B$) - temperature ($T$) phase diagram just below the critical temperature $T_C$. Soon after their first observation in MnSi,\citep{muhlbauer2009} skyrmion lattices were found in other cubic chiral magnets including Fe$_{1-x}$Co$_x$Si,\citep{munzer2010} FeGe,\citep{wilhelm2011} Cu$_2$OSeO$_3$\citep{seki2012observation,seki2012formation,adams2012} and Co-Zn-Mn alloys.\citep{tokunaga2015,karube2016} More recently, skyrmions and their lattices have been observed in polar magnets,\cite{kezsmarki2015} in thin films,\cite{tonomura2012,wilson2014} and at surfaces and interfaces of different atomic layers\cite{heinze2011}. Skyrmions are topologically stable and can be controlled with extremely small electric currents which frees the path for successful applications in novel spintronic and information storage devices.\citep{yu2012,nagaosa2013,fert2013,white2014}

In chiral magnets, the Dzyaloshinskii-Moriya interaction \cite{D,M} plays a crucial role in stabilizing the helimagnetic order and skyrmion lattices.\cite{bak1980} However, this interaction in itself cannot explain the thermodynamic stability of skyrmion lattices, which has been attributed to additional terms including thermal fluctuations,\citep{muhlbauer2009,buhrandt2013,bauer2016generic} spin exchange stiffness and/or uniaxial anisotropy.\citep{bogdanov1989,bogdanov1994,rossler2006,butenko2010,rossler2011} The specific orientation of the skyrmion lattice with respect to the crystallographic lattice can be accounted for by higher order spin-orbit coupling terms.\cite{muhlbauer2009} In order to describe this orientation dependence, it is convenient to consider the skyrmion lattice as a single domain, long-range ordered state resulting from the superposition of three helical vectors $\vec{\tau_i}$ separated by 60$^{\circ}$.\cite{adams2011,adams2012} These helices propagate in the plane perpendicular to the magnetic field and lead to the characteristic six-fold symmetric SANS patterns.\cite{muhlbauer2009} In MnSi one of the three helices preferentially aligns along the $\langle 110 \rangle$ crystallographic direction. For Cu$_2$OSeO$_3$ this orientational preference is along $\langle 100 \rangle$,\cite{adams2012,zhang2016,zhang2016prb} although some SANS measurements seem to indicate that for specific fields and temperatures the $\langle 110 \rangle$ direction is preferred.\cite{seki2012formation,makino2017}

Besides this orientational preference, patterns with twelve or more peaks have been observed by SANS and resonant x-ray scattering on Cu$_2$OSeO$_3$.\cite{langner2014,zhang2016,zhang2016prb,makino2017} These patterns indicate the coexistence of multiple skyrmion lattice domains with different orientations. In fact, such multidomain states have also been seen by Lorentz transmission electron microscopy in thin films of MnSi and Cu$_2$OSeO$_3$\citep{mochizuki2014,pollath2017} and by SANS in single crystals of Fe$_{1-x}$Co$_x$Si.\cite{munzer2010,bannenberg2016} The occurrence of these multidomain states appears for bulk Cu$_2$OSeO$_3$ to be related to the thermal history.\cite{makino2017,reim2017} In addition, it was proposed that multidomains are stabilized by magnetic fields applied along directions deviating from the major cubic axes.\citep{zhang2016prb,zhang2016} However, none of these previous studies provides a holistic view on the interrelation between the orientation of skyrmion lattices and the crystallographic lattice, the occurrence of multidomain lattices and the influence of meta-stabilities induced by different magnetic field histories for all cubic chiral magnets. 

\begin{figure}[tb]
\begin{center}
\includegraphics[width= 0.45\textwidth]{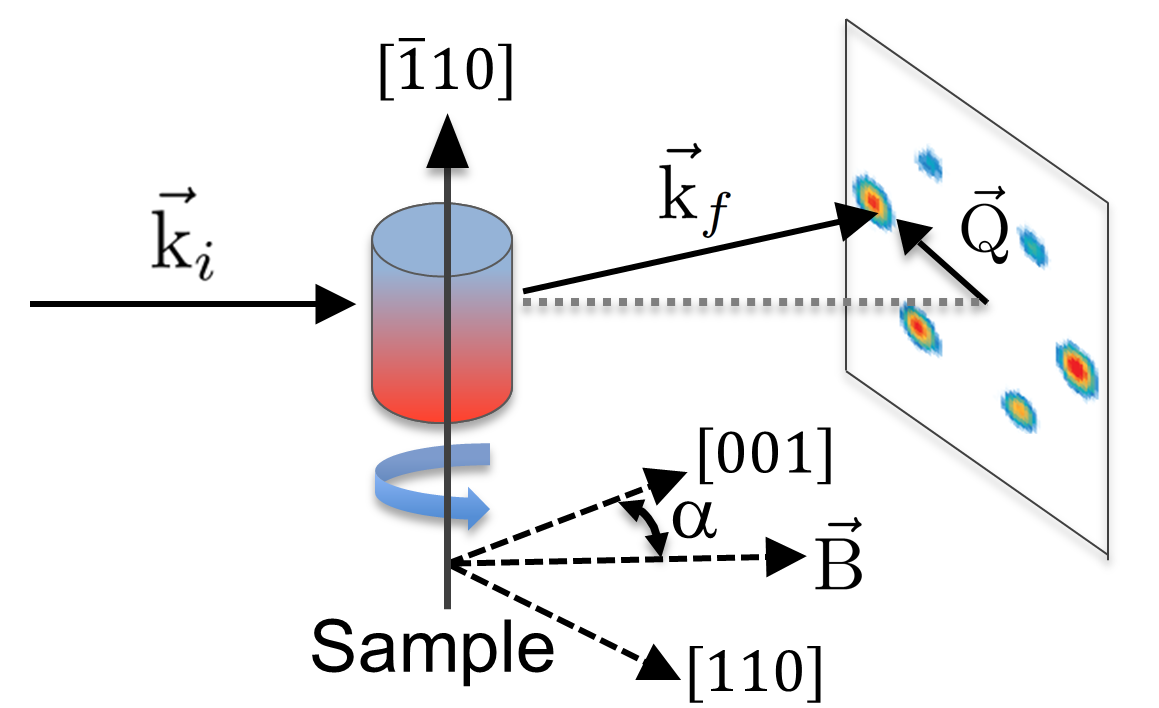}
\caption{Schematic representation of the experimental set-up. SANS measurements were performed with the magnetic field applied parallel to the incoming neutron beam with wave vector $\vec{k}_i$. The sample was aligned with the [$\bar{1}$10] crystallographic direction vertical and rotated around its vertical axis such that the magnetic field is applied along different crystallographic directions. The angle $\alpha$ is defined as the angle between the magnetic field and the $[001]$ crystallographic direction in the horizontal plane, implying that $\alpha$ = 0$^{\circ}$ corresponds to $\vec{B} \, || \,$ [001] and $\alpha$ = 90$^{\circ}$ corresponds to  $\vec{B} \, || \, [1\bar{1}0]$. The wave-vector $\vec{k}_f$ of the scattered neutron beam and the scattering vector $\vec{Q}$ are also indicated.}
\label{Exp}
\end{center}
\end{figure}

In the following we fill this gap by studying the skyrmion lattice orientation with respect to the crystallographic lattice and the occurrence of multidomain lattices in both the insulator Cu$_2$OSeO$_3$ and the archetype chiral magnet MnSi. For this purpose, we performed SANS measurements in a way so far not considered. As schematically shown in Fig. \ref{Exp}, the measurements were performed by rotating the sample around its vertical axis, and thus having the magnetic field applied along different crystallographic directions. By rotating the sample both in zero and under field, we study the effect of the magnetic field history on the orientation of the skyrmion lattice with respect to the crystallographic lattice. The results show unambiguously that the orientation of the skyrmion lattice and the occurrence of multidomain states is primarily governed by the magnetic field direction with respect to the crystallographic lattice. The orientation of the skyrmion lattice with respect to the crystallographic lattice is also influenced by the magnetic history of the sample. The latter can induce metastable orientations of the skyrmion lattice with respect to the crystallographic one that may or may not relax to their equilibrium orientation under macroscopic relaxation times. Furthermore, multidomain lattices may form when two or more equivalent crystallographic directions are favored by spin-orbit coupling and oriented perpendicular to the magnetic field.

\begin{figure*}[tb]
\begin{center}
\includegraphics[width= 1 \textwidth]{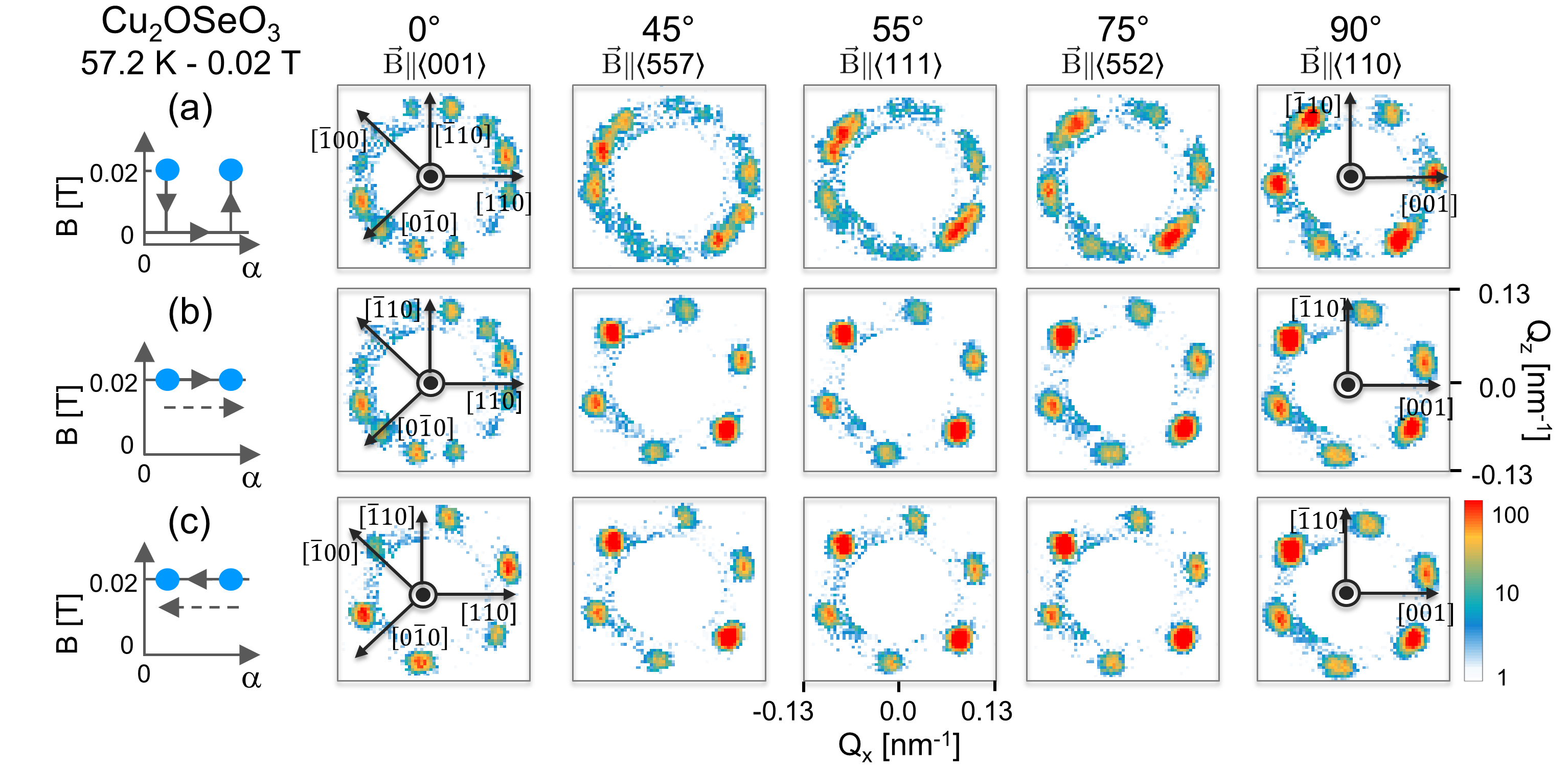}
\caption{SANS patterns showing how the skyrmion lattice orients with respect to the crystallographic lattice for different magnetic field orientation for Cu$_2$OSeO$_3$ at $T$ = 57.2~K and $B$ = 0.02~T. The field was applied along different crystallographic directions by rotating the sample in steps of 5$^\circ$ from $\alpha$ = 0$^{\circ}$ ($\vec{B} \, || \,$ [001]) to $\alpha$ = 90$^{\circ}$ ($\vec{B} \, || \, [1\bar{1}0]$) with (a) the field switched off during rotations, (b) the field continuously on, and (c) rotating the sample backwards in steps of -5$^{\circ}$ from $\alpha$ = 90$^{\circ}$ to 0$^{\circ}$ under field.}
\label{Cu2OSeO3_rot}
\end{center}
\end{figure*}

\section{Experimental}
The SANS measurements on Cu$_2$OSeO$_3$ were performed on a single crystal with dimensions of about $5\times3\times3$~mm$^3$ grown by chemical vapor transport. The sample was aligned with the [1$\bar{1}$0] crystallographic direction vertical within 3$^{\circ}$. The monochromatic SANS instrument PA20 of the Laboratoire L\'{e}on Brillouin was used with a wavelength of $\lambda$ = 0.6~nm, $\Delta\lambda$/$\lambda$ = 0.12 and the detector placed 12.7~m from the sample. The $^3$He XY multidetector is made of 128$\times$128 pixels of 5$\times$5~mm$^2$.  The magnetic field was applied parallel to the incoming neutron beam designated by its wave-vector $\vec{k}_i$ using an Oxford Instruments horizontal field cryomagnet. 

The MnSi sample is a cubic single crystal with dimensions of about $5\times5\times5$~mm$^3$ and was already used in a previous experiment.\cite{pappas2017} The MnSi sample was also aligned with the [1$\bar{1}$0] crystallographic direction vertical within 5$^{\circ}$. The SANS measurements were performed at the time-of-flight instrument LARMOR of the ISIS neutron source using neutrons with wavelengths of 0.09~nm$\, \leq \lambda  \leq \,$1.25~nm. The sample was placed 4.4~m from the detector that consists of eighty 8~mm wide $^3$He tubes. The magnetic field was applied with a three-dimensional (3D) 2T vector cryomagnet along the incoming neutron beam.\footnote{The field homogeneity of the 3D cryomagnet, defined as `the maximum field error over a 10 mm diameter spherical volume', is determined by the manufacturer as 0.11 \% of the applied magnetic field.} In particular, we eliminated the residual field of the cryomagnet by warming up the entire cryomagnet before the experiment.\footnote{We mounted a field probe on the window of the cryostat to check directly the residual field taking into account its decay with distance. This field probe confirmed the absence of a sizeable residual field. In addition, we made magnetic field scans from negative to positive fields to evaluate the residual field. As the plot of the integrated scattered intensity vs magnetic field was symmetric and centered around zero, a sizable residual field was absent.} 

All SANS patterns are normalized to standard monitor counts and background corrected using a high temperature measurement at 60~K for MnSi and 70~K for Cu$_2$OSeO$_3$.  

\begin{figure*}[tb]
\begin{center}
\includegraphics[width= 1 \textwidth]{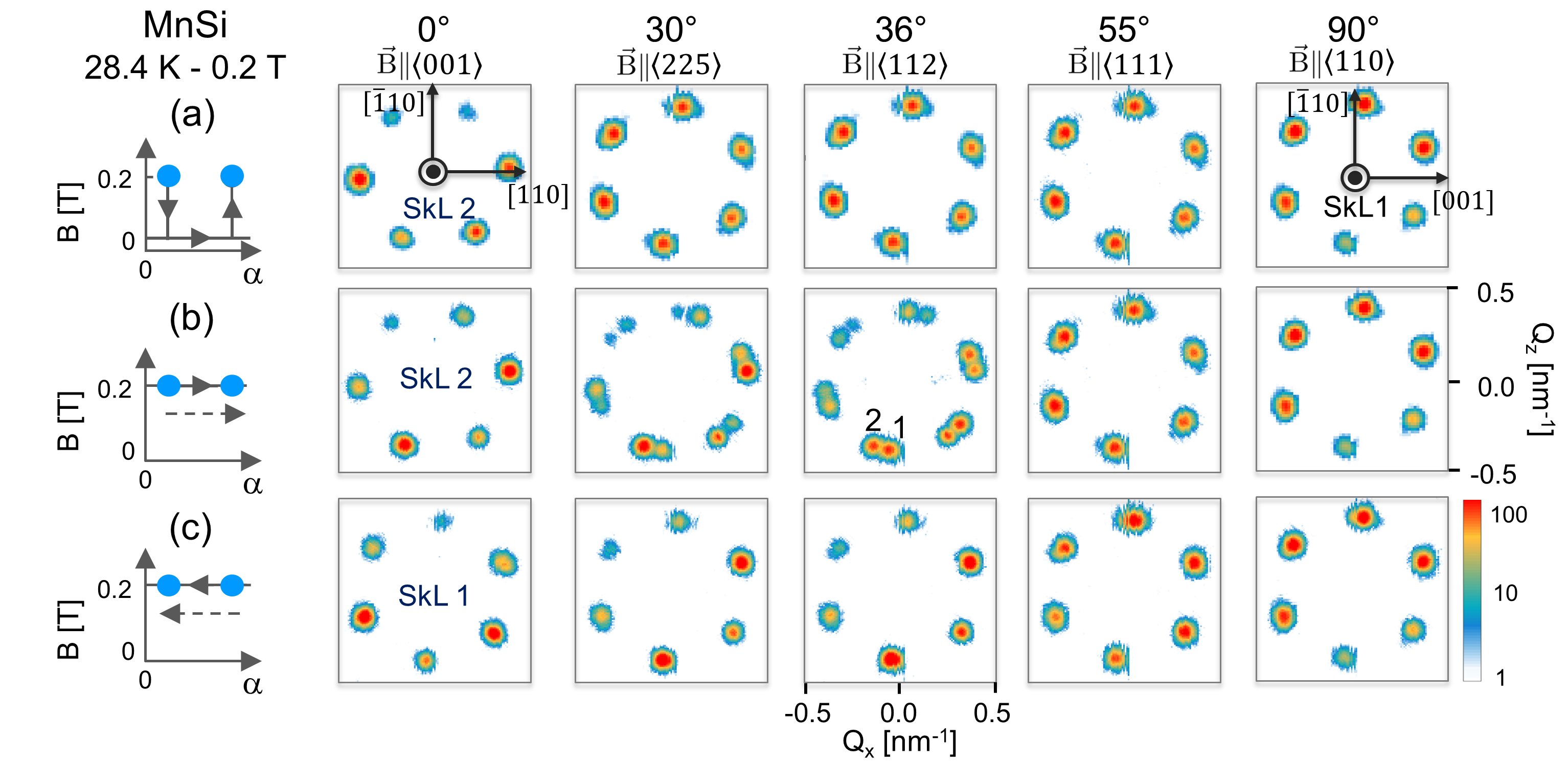}
\caption{SANS patterns showing how the skyrmion lattice orients with respect to the crystallographic lattice for different magnetic field orientation for MnSi at $T$ = 28.4~K and $B$ = 0.2~T. The field was applied along different crystallographic directions (a) after zero field cooling, (b) by rotating the sample in steps of 2$^\circ$ from $\alpha$ = 0$^{\circ}$ ($\vec{B} \, || \,$ [001]) to $\alpha$ = 90$^{\circ}$ ($\vec{B} \, || \, [1\bar{1}0]$) while keeping the field on during the rotations and (c) by rotating the sample backwards under field from $\alpha$ = 90$^{\circ}$ in steps of -2$^{\circ}$  to $\alpha$ = 0$^{\circ}$.}
\label{MnSi_rot}
\end{center}
\end{figure*}

\begin{figure*}[tb]
\begin{center}
\includegraphics[width= 1 \textwidth]{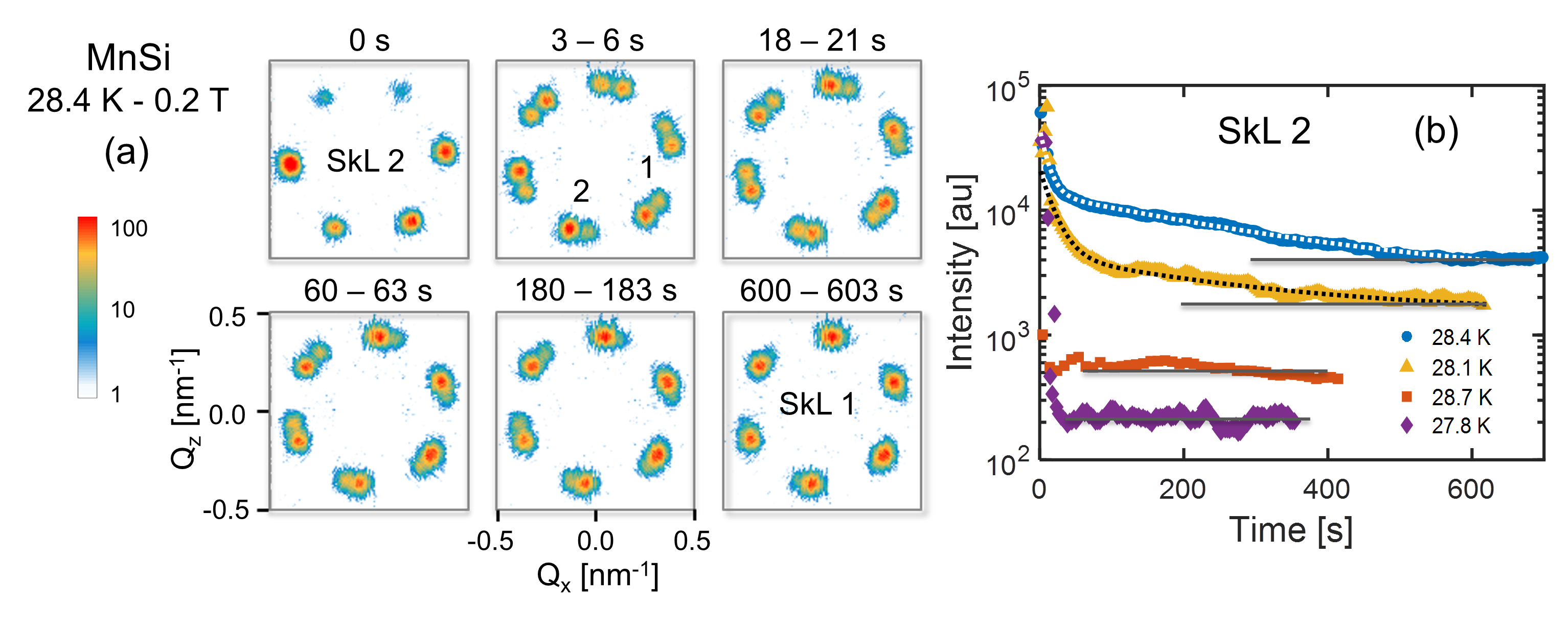}
\caption{Time-dependent relaxation of the skyrmion lattice in MnSi when the sample was rotated within 5 s from $\alpha$ = 0$^{\circ}$ ($\vec{B} \, || \, \langle 100 \rangle$) to $\alpha$ = 55$^{\circ}$ ($\vec{B} \, || \, \langle 111 \rangle$) at $B$ = 0.2~T. (a) Selection of SANS patterns obtained as a function of time at $T$ = 28.4~K. All SANS patterns can be found in Supplementary Movie 4.\cite{footnote}  (b) The total intensity of SkL 2 as a function of time for several temperatures indicated. The dotted white line for $T$ = 28.1 K and the dotted black line for $T$ = 28.4~K indicate fits of Eq.\ref{exp} to the data.  The horizontal grey-lines serve as guides to the eyes and represent the actual baseline intensity. }
\label{MnSi_event}
\end{center}
\end{figure*}

\section{Results}
\subsection{Cu$_2$OSeO$_3$}
We first consider the case of Cu$_2$OSeO$_3$ and display in Fig. \ref{Cu2OSeO3_rot}(a) and Supplemental Movie 1\cite{footnote} a series of SANS patterns measured at $T$ = 57.4~K and $B$ = 0.02~T.  The field was applied during the measurements and switched off when the sample was rotated in steps of 5$^{\circ}$ from $\alpha$ = 0$^{\circ}$ to 90$^{\circ}$. The angle $\alpha$ is defined as the angle between the magnetic field and the [001] crystallographic axis in the horizontal plane, implying that $\alpha$ = 0$^{\circ}$ corresponds to $\vec{B} \, || \,$ [001] and $\alpha$ = 90$^{\circ}$ corresponds to  $\vec{B} \, || \, [1\bar{1}0]$. 

At $\alpha$ = 0$^{\circ}$, twelve Bragg peaks are found corresponding to two six-peak patterns rotated 30$^{\circ}$ from each other. This indicates the stabilization of two distinct skyrmion lattice domains with different orientations with respect to the crystallographic lattice. One of these domains has one set of two peaks aligned along [100]  while the other domain has two peaks along the [010] direction. The intensity differences between the peaks are most likely due to a small misalignment of the single crystal and/or demagnetization effects that could also cause small differences in domain populations.

When the sample is rotated and $\alpha$ increases, the [100] and [010] directions are no longer perpendicular to the field and the skyrmion lattices can no longer orient along either of these directions. For small values of $\alpha$, the twelve-peak patterns persist implying that the coexistence of the two skyrmion domains remains energetically favorable. For larger values of $\alpha$ around 45$^{\circ}$, the patterns seem to indicate 18 peaks. Unfortunately, the resolution of the measurement is not sufficient to unambiguously confirm the existence of these 18 peaks, which would suggest the coexistence of three skyrmion lattice domains as reported elsewhere.\cite{zhang2016} 

When $\alpha$ is further increased, the third [001] direction approaches the direction perpendicular to the field and the domains gradually merge into one. At $\alpha$ = 90$^{\circ}$ only six peaks are visible and among them two are aligned along the [001] direction. At this position, the lattice has rotated by 15$^{\circ}$ with respect to each of the two lattices seen at $\alpha$ = 0$^{\circ}$.\footnote{The scattering patterns of this rotation scan are equivalent to those obtained after zero field cooling the sample.}

A similar rotation scan was performed but this time with the field on during the rotation of the sample. The  results, displayed in Fig. \ref{Cu2OSeO3_rot}(b) and Supplemental Movie 2,\cite{footnote}  show two important differences with respect to the previous case. During this rotation, the scattering from one of the two skyrmion lattice domains is suppressed, whereas the other one is enhanced. Furthermore, the SkL does not reorient for any of the field directions including at $\alpha$ = 90$^{\circ}$. At this angle the favorable [001] direction is perpendicular to the field, but the skyrmion lattice is oriented 15$^{\circ}$ away from it and is pinned by the field to the position it had at $\alpha$ = 0$^{\circ}$. This six fold symmetry persists when the sample is rotated back from $\alpha$ = 90$^{\circ}$ to 0$^{\circ}$ under magnetic field. As shown in Fig. \ref{Cu2OSeO3_rot}(c), the twelve fold symmetry is still not recovered at $\alpha$ = 0$^{\circ}$ where only one skyrmion lattice domain is found. These measurements show that the orientation of skyrmion lattices and the stabilization of multiple domains  are strongly influenced by the history of the applied magnetic field direction within the equilibrium skyrmion phase. 

\subsection{MnSi}
We now consider the case of MnSi and display in Fig. \ref{MnSi_rot}(a) a series of SANS patterns measured at $T$ = 28.4~K and $B$ = 0.2~T after zero-field cooling. For $\alpha$ = 0$^{\circ}$, i.e. for $\vec{B} \, || \, [001]$, two $\langle 110 \rangle$ directions ([110] and $[\bar{1}10]$), which are in MnSi preferred by spin-orbit coupling, are perpendicular to the magnetic field. In contrast to Cu$_2$OSeO$_3$ we do not observe twelve peaks, but only six, originating from a single SkL orientation along [110]. Six peaks are also seen for $\alpha$ $>$ 0$^{\circ}$ but this time the SkL aligns along the [$\bar{1}10$] direction. This orientation, which we name SkL 1, has a scattering pattern that is 30$^{\circ}$ rotated with respect to the one at $\alpha$ = 0$^{\circ}$ which we name SkL 2.

The MnSi sample was also rotated from $\alpha$ = 0$^{\circ}$ to 90$^{\circ}$ with a magnetic field of 0.2~T on during the measurements and the rotation of the sample. The SANS patterns were recorded every 2$^{\circ}$ and are presented in Supplemental Movie 3\cite{footnote}  with a selection given in Fig. \ref{MnSi_rot}(b). For 0 $<$ $\alpha$ $\leq$ 28$^{\circ}$, the patterns correspond to SkL 2. The patterns qualitatively change for $\alpha$ $>$ 28$^{\circ}$ and show a superposition of SkL 2 and SkL 1. By further increasing $\alpha$, SkL 2 decreases in intensity and totally vanishes at $\alpha$ $\approx$ 45$^{\circ}$, while at the same time SkL 1 becomes more prominent. As for Cu$_2$OSeO$_3$, the patterns remain the same when the sample is rotated under field back to $\alpha$ = 0$^{\circ}$. SkL 2 is still not recovered after waiting 30 min at $\alpha$ = 0$^{\circ}$. These results indicate that in both Cu$_2$OSeO$_3$ and MnSi the orientation of the SkL is determined by the magnetic field orientation with respect to crystallographic lattice as well as by the history of the magnetic field (direction) of the sample. They also demonstrate that it is possible to induce metastable skyrmion states in both Cu$_2$OSeO$_3$ and MnSi within the thermodynamic equilibrium skyrmion phase.

In order to follow the re-orientation from SkL 2 to SkL 1 as a function of time, we performed fast rotations from $\alpha$ = 0$^{\circ}$ ($\vec{B} \, || \, \langle 100 \rangle$) to $\alpha$ = 55$^{\circ}$ ($\vec{B} \, || \, \langle 111 \rangle$) while keeping the field on. The SANS patterns where recorded continuously using event mode data acquisition. Subsequently, they were processed such that a pattern was obtained for every 3~s. All patterns are displayed in Supplemental Movie 4\cite{footnote}  and a selection of them is shown in Fig. \ref{MnSi_event}(a). During the first 3 seconds, SkL 1 forms and coexists with SkL 2. Afterwards, the intensity of SkL 2 drops gradually whereas the intensity of SkL 1 increases. After $t$ $\approx$ 25~s, the rate at which the intensity of SkL 2 decreases slows down considerably, and it takes almost ten minutes for SkL 2 to completely disappear. 

\begin{figure*}[tb]
\begin{center}
\includegraphics[width= 0.95\textwidth]{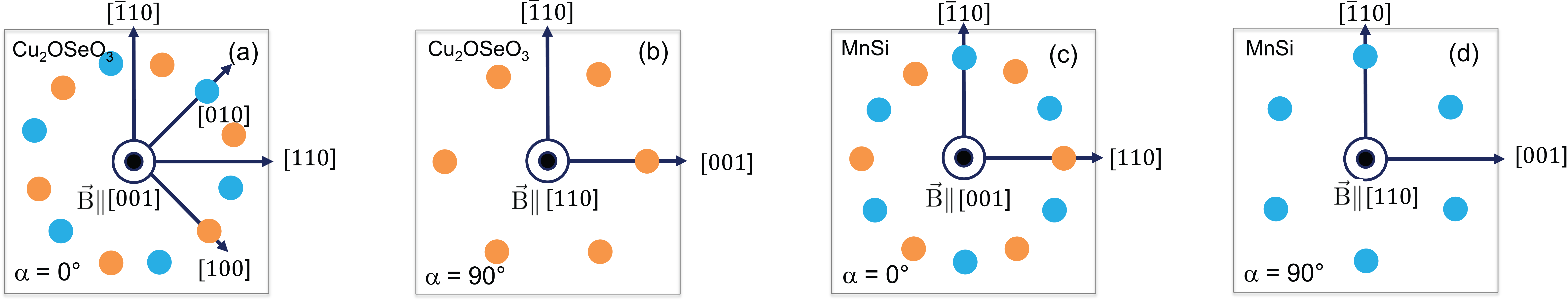}
\caption{Schematic illustration of the skyrmion lattice orientations with the minimum energy for (a) and(b) Cu$_2$OSeO$_3$, where spin-orbit coupling prefers Bragg peaks along $\langle 100 \rangle$, and  (c) and (d) MnSi, where it prefers peaks along $\langle 110 \rangle$. The illustrations correspond to the characteristic six Bragg peak scattering pattern of a skyrmion lattice domain seen by SANS. Different colors represent different domains of skyrmion lattices. The magnetic field is applied in the out of plane direction along (a) and (c) $\vec{B} \, || \, [001]$ ($\alpha$ = 0$^{\circ}$) and (b) and (d) $\vec{B} \, || \,[1\bar{1}0]$ ($\alpha$ = 90$^{\circ}$).}
\label{Model}
\end{center}
\end{figure*}

A quantitative analysis of this reorientation is provided by considering the time dependence of the total intensity of all six Bragg peaks of SkL 2. The results are displayed in Fig. \ref{MnSi_event}(b) and show a fast decrease in intensity within the first 25 seconds followed by a slow decay. Therefore, we fitted the data to a superposition of two exponentials:

\begin{equation}
I(t) = a \exp(-t/t_1)+b \exp(-t/t_2) + c.
\label{exp}
\end{equation}

\noindent The fit at $T$ = 28.4~K provides $a$ = 3.1 $\pm$ 0.1 $\times$ 10$^{4}$, $b$ = 1.1 $\pm$ 0.1 $\times$ 10$^{4}$, $c$ = 2.5 $\pm$ 0.2 $\times$ 10$^{4}$ and the two time constants $t_1$ = 11.2 $\pm$ 0.6~s and $t_2$ = 3.0 $\pm$ 0.3 $\times$ 10$^{2}$~s. As such, it shows that the relaxation is governed by two separate processes occurring on the seconds and minutes time-scales, respectively. These processes possibly reflect the movement of (topological) domain walls and/or their pinning to defects of the crystallographic lattice.\cite{pollath2017} Slow relaxations that originate from multiple processes have also been observed around the A-phase by ac magnetic susceptibility measurements in Fe$_{1-x}$Co$_x$Si.\cite{bannenberg2016squid}

The relaxation of skyrmion lattices does not change dramatically in the center of the A-phase. At a lower temperature of 28.1 K, the estimated values are $t_1$ = 18.9 $\pm$ 0.4~s and $t_2$ = 2.6 $\pm$ 0.2 $\times$ 10$^{2}$~s. Thus, the characteristic time of the slow relaxation remains almost unchanged, whereas that of the faster process is almost doubled. The relaxation times are very different near the low- and high temperature borders as shown for 27.8 and 28.7~K in Fig. \ref{MnSi_event}(b). The acceleration of the relaxation is not surprising for the high temperature border where it can be attributed to increased thermal fluctuations. The acceleration at the low temperature limit of $T$ = 27.8~K may be due to the fact that at this temperature the SkL is stable only for $\vec{B} \, || \, \langle 001 \rangle$.\cite{bauer2010} Indeed, we observe that the skyrmion lattice relaxes to the conical phase at this temperature and the SkL is thus not stable for $\vec{B} \, || \, \langle 111 \rangle$. This highlights the importance of anisotropy terms.




\section{Discussion and Conclusion}
The results presented above show that the skyrmion lattice orients along the crystallographic directions expected from spin-orbit coupling when the samples are zero field cooled. Indeed, based on a Ginzburg-Landau analysis,\cite{muhlbauer2009,munzer2010,seki2012formation} one expects from the relevant fourth and sixth order spin-orbit coupling terms $\sum_\tau (\tau_x^6+\tau_y^6+\tau_z^6)\left|\vec{m}_\tau\right|^2$ and $\sum_\tau (\tau_x^4\tau_y^2+\tau_y^4\tau_z^4+\tau_z^4\tau_x^4)\left|\vec{m}_\tau\right|^2$, with $\vec{m}_\tau$ being the Fourier transform of the local magnetization $\vec{M}(\vec{r})$, an alignment of the skyrmion lattice with one of the helical vectors $\vec{\tau} \, || \,\langle 100 \rangle$ as for MnSi, or $\vec{\tau} \, || \,\langle 110 \rangle$ as for Cu$_2$OSeO$_3$.

Multidomain lattices may form when several equivalent crystallographic directions, as preferred by spin-orbit coupling, are simultaneously perpendicular to the magnetic field. This is exemplified by the schematic drawings in Fig. \ref{Model}. Figures \ref{Model}(a) and \ref{Model}(c) show that two domains of skyrmion lattices are expected for both Cu$_2$OSeO$_3$ and MnSi for $\alpha$ = 0$^{\circ}$ and $\vec{B} \, || \,\langle 100 \rangle$, where there are two favorable $\langle 100 \rangle$ and $\langle 110 \rangle$ crystallographic directions perpendicular to the field. In the other configuration of $\alpha$ = 90$^{\circ}$ and $\vec{B} \, || \,\langle 110 \rangle$ shown in Fig. \ref{Model}(b) and \ref{Model}(d), only one $\langle 100 \rangle$ and $\langle 110 \rangle$ crystallographic direction is perpendicular to the field, and hence, only one SkL domain is expected for both chiral magnets.  A comparison with experiment shows, however, that multidomain SkLs are  nucleated only in Cu$_2$OSeO$_3$. This is in agreement with the literature, where, to our knowledge, no stable multidomain lattice has been reported for bulk single-crystal MnSi so far. 

This important difference between Cu$_2$OSeO$_3$ and MnSi may be attributed to substantial differences in the most relevant terms in the free energy Landau-Ginzburg functional that contains the Ferromagnetic exchange, Dzyaloshinskii-Moriya interaction, Zeeman energy and anisotropy/spin-orbit coupling terms.\cite{muhlbauer2009} The most obvious difference is in the coupling to the external field and the Zeeman energy, which is almost an order of magnitude stronger for MnSi than Cu$_2$OSeO$_3$. Indeed, the magnetic fields required to stabilize the skyrmion lattice phase are almost an order of magnitude stronger for MnSi than for Cu$_2$OSeO$_3$ although the volume magnetizations of both systems are very similar to each other.\cite{bauer2010,qian2016} An additional more subtle but most relevant difference is in the spin-orbit coupling that pins the skyrmion lattice to the crystallographic one. The higher order spin-orbit coupling terms seem to be significantly stronger for MnSi than for Cu$_2$OSeO$_3$. In the later, both the fourth and sixth order terms responsible for this coupling are very weak as pointed out by a previous study.\cite{seki2012formation} Consequently, multidomain lattices are stabilized in Cu$_2$OSeO$_3$ for a wide range of field directions with respect to the crystallographic lattice. A similar case has also been documented for Fe$_{1-x}$Co$_x$Si where the coupling of the skyrmion lattice to the crystallographic lattice is likely even weaker.\cite{munzer2010,bannenberg2016} If the spin-orbit coupling is stronger, as is likely the case for MnSi, a different field orientation with respect to the crystallographic lattice has a larger impact on the energy levels of the energy minima. Thus, in MnSi both the stronger Zeeman and spin orbit coupling terms in conjunction with even small sample misalignments, imperfections, or demagnetizing fields would raise the degeneracy between different and equivalent skyrmion lattice domains and thus favor a single domain configuration. We therefore conjecture that multidomain lattices should also exist in MnSi, but only within a very narrow region of field orientations with respect to the crystallographic lattice that has not been realized experimentally until now.

Our results also show that the specific history of the magnetic field (direction) has a significant impact on the SkL orientation. When rotations are performed under field, the multidomain SkL stabilized for $\vec{B} \, || \,\langle 100 \rangle$ in Cu$_2$OSeO$_3$ evolves to a single-domain SkL. Upon further rotation, this single domain does not reorient to its zero-field cooled configuration, which one may assume to be the most energetically favorable one. On the other hand, the skyrmion lattice may reorient under certain conditions for MnSi involving macroscopic relaxation times and metastable multidomain lattices. Thus, for both systems relatively large energy barriers prevent  SkLs from reorienting to their thermodynamically most favorable state. The existence
of such high energy barriers is not surprising as such a reorientation of the skyrmion lattice involves a rearrangement of the magnetic configuration over very large (macroscopic) volumes. These results thus show that it is possible to induce metastable skyrmion states in Cu$_2$OSeO$_3$ and MnSi within the thermodynamic equilibrium skyrmion phase. 

The stabilization of the multidomain SkL in Cu$_2$OSeO$_3$ has previously been attributed to magnetic field directions deviating from the major cubic axes\cite{zhang2016,zhang2016prb} or to the thermal and magnetic history.\cite{seki2012formation} Our results show that specific magnetic field histories can indeed suppress multidomain lattices. However, in contrast to previous work,\cite{zhang2016,zhang2016prb} we find that multidomain SkLs can also be stabilized when the field is applied along a major cubic axis such as the $\langle 100 \rangle$  crystallographic direction. The occurrence of multidomain lattices can thus be understood on the basis of symmetry arguments, as illustrated in Fig. \ref{Model}.

In summary, the results presented above show that the orientation of the skyrmion lattice is  governed primarily by the magnetic field direction with respect to the crystallographic lattice, but is also influenced by the magnetic history of the sample. Multidomain lattices may form when two or more equivalent crystallographic directions are favored by spin-orbit coupling and oriented perpendicular to the magnetic field. These results provide new insights into the factors that stabilize skyrmion lattices and influence their orientation.  They shed new light on the puzzle of the occurrence of multiple skyrmion lattice domains, an issue that is of general relevance to chiral magnetism. 

\begin{acknowledgments}
The authors wish to express their gratitude to the ISIS and LLB technical support staff for their assistance and are grateful for the kind help of C. Decorse with aligning the single crystals. The experiments at the ISIS Pulsed Neutron and Muon Source were supported by a beamtime allocation from the Science and Technology Facilities Council and the Netherlands Organization for Scientific Research (NWO). The work of L.J.B and C.P. is financially supported by The Netherlands Organization for Scientific Research through Project No. 721.012.102. F.Q. thanks the China Scholarship Council for financial support. D.L.S. and T.A.L. acknowledge support from  the U.S. Department of Energy, Office of Basic Energy Sciences, Division of Materials Sciences and Engineering under Contract No DE-AC02-07CH11358. 
\end{acknowledgments}

\bibliography{MnSi_Cu2OSeO3_Rotation}

\begin{thebibliography}{46}%
\makeatletter
\providecommand \@ifxundefined [1]{%
 \@ifx{#1\undefined}
}%
\providecommand \@ifnum [1]{%
 \ifnum #1\expandafter \@firstoftwo
 \else \expandafter \@secondoftwo
 \fi
}%
\providecommand \@ifx [1]{%
 \ifx #1\expandafter \@firstoftwo
 \else \expandafter \@secondoftwo
 \fi
}%
\providecommand \natexlab [1]{#1}%
\providecommand \enquote  [1]{``#1''}%
\providecommand \bibnamefont  [1]{#1}%
\providecommand \bibfnamefont [1]{#1}%
\providecommand \citenamefont [1]{#1}%
\providecommand \href@noop [0]{\@secondoftwo}%
\providecommand \href [0]{\begingroup \@sanitize@url \@href}%
\providecommand \@href[1]{\@@startlink{#1}\@@href}%
\providecommand \@@href[1]{\endgroup#1\@@endlink}%
\providecommand \@sanitize@url [0]{\catcode `\\12\catcode `\$12\catcode
  `\&12\catcode `\#12\catcode `\^12\catcode `\_12\catcode `\%12\relax}%
\providecommand \@@startlink[1]{}%
\providecommand \@@endlink[0]{}%
\providecommand \url  [0]{\begingroup\@sanitize@url \@url }%
\providecommand \@url [1]{\endgroup\@href {#1}{\urlprefix }}%
\providecommand \urlprefix  [0]{URL }%
\providecommand \Eprint [0]{\href }%
\providecommand \doibase [0]{http://dx.doi.org/}%
\providecommand \selectlanguage [0]{\@gobble}%
\providecommand \bibinfo  [0]{\@secondoftwo}%
\providecommand \bibfield  [0]{\@secondoftwo}%
\providecommand \translation [1]{[#1]}%
\providecommand \BibitemOpen [0]{}%
\providecommand \bibitemStop [0]{}%
\providecommand \bibitemNoStop [0]{.\EOS\space}%
\providecommand \EOS [0]{\spacefactor3000\relax}%
\providecommand \BibitemShut  [1]{\csname bibitem#1\endcsname}%
\let\auto@bib@innerbib\@empty
\bibitem [{\citenamefont {M{\"u}hlbauer}\ \emph {et~al.}(2009)\citenamefont
  {M{\"u}hlbauer}, \citenamefont {Binz}, \citenamefont {Jonietz}, \citenamefont
  {Pfleiderer}, \citenamefont {Rosch}, \citenamefont {Neubauer}, \citenamefont
  {Georgii},\ and\ \citenamefont {B{\"o}ni}}]{muhlbauer2009}%
  \BibitemOpen
  \bibfield  {author} {\bibinfo {author} {\bibfnamefont {S.}~\bibnamefont
  {M{\"u}hlbauer}}, \bibinfo {author} {\bibfnamefont {B.}~\bibnamefont {Binz}},
  \bibinfo {author} {\bibfnamefont {F.}~\bibnamefont {Jonietz}}, \bibinfo
  {author} {\bibfnamefont {C.}~\bibnamefont {Pfleiderer}}, \bibinfo {author}
  {\bibfnamefont {A.}~\bibnamefont {Rosch}}, \bibinfo {author} {\bibfnamefont
  {A.}~\bibnamefont {Neubauer}}, \bibinfo {author} {\bibfnamefont
  {R.}~\bibnamefont {Georgii}}, \ and\ \bibinfo {author} {\bibfnamefont
  {P.}~\bibnamefont {B{\"o}ni}},\ }\href@noop {} {\bibfield  {journal}
  {\bibinfo  {journal} {Science}\ }\textbf {\bibinfo {volume} {323}},\ \bibinfo
  {pages} {915} (\bibinfo {year} {2009})}\BibitemShut {NoStop}%
\bibitem [{\citenamefont {Fert}\ \emph {et~al.}(2013)\citenamefont {Fert},
  \citenamefont {Cros},\ and\ \citenamefont {Sampaio}}]{fert2013}%
  \BibitemOpen
  \bibfield  {author} {\bibinfo {author} {\bibfnamefont {A.}~\bibnamefont
  {Fert}}, \bibinfo {author} {\bibfnamefont {V.}~\bibnamefont {Cros}}, \ and\
  \bibinfo {author} {\bibfnamefont {J.}~\bibnamefont {Sampaio}},\ }\href@noop
  {} {\bibfield  {journal} {\bibinfo  {journal} {Nature Nanotechnology}\
  }\textbf {\bibinfo {volume} {8}},\ \bibinfo {pages} {152} (\bibinfo {year}
  {2013})}\BibitemShut {NoStop}%
\bibitem [{\citenamefont {Nagaosa}\ and\ \citenamefont
  {Tokura}(2013)}]{nagaosa2013}%
  \BibitemOpen
  \bibfield  {author} {\bibinfo {author} {\bibfnamefont {N.}~\bibnamefont
  {Nagaosa}}\ and\ \bibinfo {author} {\bibfnamefont {Y.}~\bibnamefont
  {Tokura}},\ }\href@noop {} {\bibfield  {journal} {\bibinfo  {journal} {Nature
  Nanotechnology}\ }\textbf {\bibinfo {volume} {8}},\ \bibinfo {pages} {899}
  (\bibinfo {year} {2013})}\BibitemShut {NoStop}%
\bibitem [{\citenamefont {Milde}\ \emph {et~al.}(2013)\citenamefont {Milde},
  \citenamefont {K{\"o}hler}, \citenamefont {Seidel}, \citenamefont {Eng},
  \citenamefont {Bauer}, \citenamefont {Chacon}, \citenamefont {Kindervater},
  \citenamefont {M{\"u}hlbauer}, \citenamefont {Pfleiderer}, \citenamefont
  {Buhrandt} \emph {et~al.}}]{milde2013}%
  \BibitemOpen
  \bibfield  {author} {\bibinfo {author} {\bibfnamefont {P.}~\bibnamefont
  {Milde}}, \bibinfo {author} {\bibfnamefont {D.}~\bibnamefont {K{\"o}hler}},
  \bibinfo {author} {\bibfnamefont {J.}~\bibnamefont {Seidel}}, \bibinfo
  {author} {\bibfnamefont {L.}~\bibnamefont {Eng}}, \bibinfo {author}
  {\bibfnamefont {A.}~\bibnamefont {Bauer}}, \bibinfo {author} {\bibfnamefont
  {A.}~\bibnamefont {Chacon}}, \bibinfo {author} {\bibfnamefont
  {J.}~\bibnamefont {Kindervater}}, \bibinfo {author} {\bibfnamefont
  {S.}~\bibnamefont {M{\"u}hlbauer}}, \bibinfo {author} {\bibfnamefont
  {C.}~\bibnamefont {Pfleiderer}}, \bibinfo {author} {\bibfnamefont
  {S.}~\bibnamefont {Buhrandt}},  \emph {et~al.},\ }\href@noop {} {\bibfield
  {journal} {\bibinfo  {journal} {Science}\ }\textbf {\bibinfo {volume}
  {340}},\ \bibinfo {pages} {1076} (\bibinfo {year} {2013})}\BibitemShut
  {NoStop}%
\bibitem [{\citenamefont {Romming}\ \emph {et~al.}(2013)\citenamefont
  {Romming}, \citenamefont {Hanneken}, \citenamefont {Menzel}, \citenamefont
  {Bickel}, \citenamefont {Wolter}, \citenamefont {von Bergmann}, \citenamefont
  {Kubetzka},\ and\ \citenamefont {Wiesendanger}}]{romming2013}%
  \BibitemOpen
  \bibfield  {author} {\bibinfo {author} {\bibfnamefont {N.}~\bibnamefont
  {Romming}}, \bibinfo {author} {\bibfnamefont {C.}~\bibnamefont {Hanneken}},
  \bibinfo {author} {\bibfnamefont {M.}~\bibnamefont {Menzel}}, \bibinfo
  {author} {\bibfnamefont {J.~E.}\ \bibnamefont {Bickel}}, \bibinfo {author}
  {\bibfnamefont {B.}~\bibnamefont {Wolter}}, \bibinfo {author} {\bibfnamefont
  {K.}~\bibnamefont {von Bergmann}}, \bibinfo {author} {\bibfnamefont
  {A.}~\bibnamefont {Kubetzka}}, \ and\ \bibinfo {author} {\bibfnamefont
  {R.}~\bibnamefont {Wiesendanger}},\ }\href@noop {} {\bibfield  {journal}
  {\bibinfo  {journal} {Science}\ }\textbf {\bibinfo {volume} {341}},\ \bibinfo
  {pages} {636} (\bibinfo {year} {2013})}\BibitemShut {NoStop}%
\bibitem [{\citenamefont {Seki}\ and\ \citenamefont
  {Mochizuki}(2015)}]{seki2015skyrmions}%
  \BibitemOpen
  \bibfield  {author} {\bibinfo {author} {\bibfnamefont {S.}~\bibnamefont
  {Seki}}\ and\ \bibinfo {author} {\bibfnamefont {M.}~\bibnamefont
  {Mochizuki}},\ }\href@noop {} {\emph {\bibinfo {title} {Skyrmions in Magnetic
  Materials}}}\ (\bibinfo  {publisher} {Springer},\ \bibinfo {year}
  {2015})\BibitemShut {NoStop}%
\bibitem [{\citenamefont {Bauer}\ and\ \citenamefont
  {Pfleiderer}(2016)}]{bauer2016generic}%
  \BibitemOpen
  \bibfield  {author} {\bibinfo {author} {\bibfnamefont {A.}~\bibnamefont
  {Bauer}}\ and\ \bibinfo {author} {\bibfnamefont {C.}~\bibnamefont
  {Pfleiderer}},\ }in\ \href@noop {} {\emph {\bibinfo {booktitle} {Topological
  Structures in Ferroic Materials}}}\ (\bibinfo  {publisher} {Springer},\
  \bibinfo {year} {2016})\ pp.\ \bibinfo {pages} {1--28}\BibitemShut {NoStop}%
\bibitem [{\citenamefont {M{\"u}nzer}\ \emph {et~al.}(2010)\citenamefont
  {M{\"u}nzer}, \citenamefont {Neubauer}, \citenamefont {Adams}, \citenamefont
  {M{\"u}hlbauer}, \citenamefont {Franz}, \citenamefont {Jonietz},
  \citenamefont {Georgii}, \citenamefont {B{\"o}ni}, \citenamefont {Pedersen},
  \citenamefont {Schmidt} \emph {et~al.}}]{munzer2010}%
  \BibitemOpen
  \bibfield  {author} {\bibinfo {author} {\bibfnamefont {W.}~\bibnamefont
  {M{\"u}nzer}}, \bibinfo {author} {\bibfnamefont {A.}~\bibnamefont
  {Neubauer}}, \bibinfo {author} {\bibfnamefont {T.}~\bibnamefont {Adams}},
  \bibinfo {author} {\bibfnamefont {S.}~\bibnamefont {M{\"u}hlbauer}}, \bibinfo
  {author} {\bibfnamefont {C.}~\bibnamefont {Franz}}, \bibinfo {author}
  {\bibfnamefont {F.}~\bibnamefont {Jonietz}}, \bibinfo {author} {\bibfnamefont
  {R.}~\bibnamefont {Georgii}}, \bibinfo {author} {\bibfnamefont
  {P.}~\bibnamefont {B{\"o}ni}}, \bibinfo {author} {\bibfnamefont
  {B.}~\bibnamefont {Pedersen}}, \bibinfo {author} {\bibfnamefont
  {M.}~\bibnamefont {Schmidt}},  \emph {et~al.},\ }\href@noop {} {\bibfield
  {journal} {\bibinfo  {journal} {Physical Review B}\ }\textbf {\bibinfo
  {volume} {81}},\ \bibinfo {pages} {041203} (\bibinfo {year}
  {2010})}\BibitemShut {NoStop}%
\bibitem [{\citenamefont {Wilhelm}\ \emph {et~al.}(2011)\citenamefont
  {Wilhelm}, \citenamefont {Baenitz}, \citenamefont {Schmidt}, \citenamefont
  {R{\"o}{\ss}ler}, \citenamefont {Leonov},\ and\ \citenamefont
  {Bogdanov}}]{wilhelm2011}%
  \BibitemOpen
  \bibfield  {author} {\bibinfo {author} {\bibfnamefont {H.}~\bibnamefont
  {Wilhelm}}, \bibinfo {author} {\bibfnamefont {M.}~\bibnamefont {Baenitz}},
  \bibinfo {author} {\bibfnamefont {M.}~\bibnamefont {Schmidt}}, \bibinfo
  {author} {\bibfnamefont {U.~K.}\ \bibnamefont {R{\"o}{\ss}ler}}, \bibinfo
  {author} {\bibfnamefont {A.~A.}\ \bibnamefont {Leonov}}, \ and\ \bibinfo
  {author} {\bibfnamefont {A.~N.}\ \bibnamefont {Bogdanov}},\ }\href@noop {}
  {\bibfield  {journal} {\bibinfo  {journal} {Physical Review Letters}\
  }\textbf {\bibinfo {volume} {107}},\ \bibinfo {pages} {127203} (\bibinfo
  {year} {2011})}\BibitemShut {NoStop}%
\bibitem [{\citenamefont {Seki}\ \emph
  {et~al.}(2012{\natexlab{a}})\citenamefont {Seki}, \citenamefont {Yu},
  \citenamefont {Ishiwata},\ and\ \citenamefont
  {Tokura}}]{seki2012observation}%
  \BibitemOpen
  \bibfield  {author} {\bibinfo {author} {\bibfnamefont {S.}~\bibnamefont
  {Seki}}, \bibinfo {author} {\bibfnamefont {X.}~\bibnamefont {Yu}}, \bibinfo
  {author} {\bibfnamefont {S.}~\bibnamefont {Ishiwata}}, \ and\ \bibinfo
  {author} {\bibfnamefont {Y.}~\bibnamefont {Tokura}},\ }\href@noop {}
  {\bibfield  {journal} {\bibinfo  {journal} {Science}\ }\textbf {\bibinfo
  {volume} {336}},\ \bibinfo {pages} {198} (\bibinfo {year}
  {2012}{\natexlab{a}})}\BibitemShut {NoStop}%
\bibitem [{\citenamefont {Seki}\ \emph
  {et~al.}(2012{\natexlab{b}})\citenamefont {Seki}, \citenamefont {Kim},
  \citenamefont {Inosov}, \citenamefont {Georgii}, \citenamefont {Keimer},
  \citenamefont {Ishiwata},\ and\ \citenamefont {Tokura}}]{seki2012formation}%
  \BibitemOpen
  \bibfield  {author} {\bibinfo {author} {\bibfnamefont {S.}~\bibnamefont
  {Seki}}, \bibinfo {author} {\bibfnamefont {J.-H.}\ \bibnamefont {Kim}},
  \bibinfo {author} {\bibfnamefont {D.~S.}\ \bibnamefont {Inosov}}, \bibinfo
  {author} {\bibfnamefont {R.}~\bibnamefont {Georgii}}, \bibinfo {author}
  {\bibfnamefont {B.}~\bibnamefont {Keimer}}, \bibinfo {author} {\bibfnamefont
  {S.}~\bibnamefont {Ishiwata}}, \ and\ \bibinfo {author} {\bibfnamefont
  {Y.}~\bibnamefont {Tokura}},\ }\href@noop {} {\bibfield  {journal} {\bibinfo
  {journal} {Physical Review B}\ }\textbf {\bibinfo {volume} {85}},\ \bibinfo
  {pages} {220406} (\bibinfo {year} {2012}{\natexlab{b}})}\BibitemShut
  {NoStop}%
\bibitem [{\citenamefont {Adams}\ \emph {et~al.}(2012)\citenamefont {Adams},
  \citenamefont {Chacon}, \citenamefont {Wagner}, \citenamefont {Bauer},
  \citenamefont {Brandl}, \citenamefont {Pedersen}, \citenamefont {Berger},
  \citenamefont {Lemmens},\ and\ \citenamefont {Pfleiderer}}]{adams2012}%
  \BibitemOpen
  \bibfield  {author} {\bibinfo {author} {\bibfnamefont {T.}~\bibnamefont
  {Adams}}, \bibinfo {author} {\bibfnamefont {A.}~\bibnamefont {Chacon}},
  \bibinfo {author} {\bibfnamefont {M.}~\bibnamefont {Wagner}}, \bibinfo
  {author} {\bibfnamefont {A.}~\bibnamefont {Bauer}}, \bibinfo {author}
  {\bibfnamefont {G.}~\bibnamefont {Brandl}}, \bibinfo {author} {\bibfnamefont
  {B.}~\bibnamefont {Pedersen}}, \bibinfo {author} {\bibfnamefont
  {H.}~\bibnamefont {Berger}}, \bibinfo {author} {\bibfnamefont
  {P.}~\bibnamefont {Lemmens}}, \ and\ \bibinfo {author} {\bibfnamefont
  {C.}~\bibnamefont {Pfleiderer}},\ }\href@noop {} {\bibfield  {journal}
  {\bibinfo  {journal} {Physical Review Letters}\ }\textbf {\bibinfo {volume}
  {108}},\ \bibinfo {pages} {237204} (\bibinfo {year} {2012})}\BibitemShut
  {NoStop}%
\bibitem [{\citenamefont {Tokunaga}\ \emph {et~al.}(2015)\citenamefont
  {Tokunaga}, \citenamefont {Yu}, \citenamefont {White}, \citenamefont
  {R{\o}nnow}, \citenamefont {Morikawa}, \citenamefont {Taguchi},\ and\
  \citenamefont {Tokura}}]{tokunaga2015}%
  \BibitemOpen
  \bibfield  {author} {\bibinfo {author} {\bibfnamefont {Y.}~\bibnamefont
  {Tokunaga}}, \bibinfo {author} {\bibfnamefont {X.~Z.}\ \bibnamefont {Yu}},
  \bibinfo {author} {\bibfnamefont {J.~S.}\ \bibnamefont {White}}, \bibinfo
  {author} {\bibfnamefont {H.~M.}\ \bibnamefont {R{\o}nnow}}, \bibinfo {author}
  {\bibfnamefont {D.}~\bibnamefont {Morikawa}}, \bibinfo {author}
  {\bibfnamefont {Y.}~\bibnamefont {Taguchi}}, \ and\ \bibinfo {author}
  {\bibfnamefont {Y.}~\bibnamefont {Tokura}},\ }\href@noop {} {\bibfield
  {journal} {\bibinfo  {journal} {Nature Communications}\ }\textbf {\bibinfo
  {volume} {6}},\ \bibinfo {pages} {7638} (\bibinfo {year} {2015})}\BibitemShut
  {NoStop}%
\bibitem [{\citenamefont {Karube}\ \emph {et~al.}(2016)\citenamefont {Karube},
  \citenamefont {White}, \citenamefont {Reynolds}, \citenamefont {Gavilano},
  \citenamefont {Oike}, \citenamefont {Kikkawa}, \citenamefont {Kagawa},
  \citenamefont {Tokunaga}, \citenamefont {R{\o}nnow},\ and\ \citenamefont
  {Tokura}}]{karube2016}%
  \BibitemOpen
  \bibfield  {author} {\bibinfo {author} {\bibfnamefont {K.}~\bibnamefont
  {Karube}}, \bibinfo {author} {\bibfnamefont {J.~S.}\ \bibnamefont {White}},
  \bibinfo {author} {\bibfnamefont {N.}~\bibnamefont {Reynolds}}, \bibinfo
  {author} {\bibfnamefont {J.~L.}\ \bibnamefont {Gavilano}}, \bibinfo {author}
  {\bibfnamefont {H.}~\bibnamefont {Oike}}, \bibinfo {author} {\bibfnamefont
  {A.}~\bibnamefont {Kikkawa}}, \bibinfo {author} {\bibfnamefont
  {F.}~\bibnamefont {Kagawa}}, \bibinfo {author} {\bibfnamefont
  {Y.}~\bibnamefont {Tokunaga}}, \bibinfo {author} {\bibfnamefont {H.~M.}\
  \bibnamefont {R{\o}nnow}}, \ and\ \bibinfo {author} {\bibfnamefont
  {Y.~e.~a.}\ \bibnamefont {Tokura}},\ }\href@noop {} {\bibfield  {journal}
  {\bibinfo  {journal} {Nature Materials}\ }\textbf {\bibinfo {volume} {14}},\
  \bibinfo {pages} {1116} (\bibinfo {year} {2016})}\BibitemShut {NoStop}%
\bibitem [{\citenamefont {K{\'e}zsm{\'a}rki}\ \emph {et~al.}(2015)\citenamefont
  {K{\'e}zsm{\'a}rki}, \citenamefont {Bord{\'a}cs}, \citenamefont {Milde},
  \citenamefont {Neuber}, \citenamefont {Eng}, \citenamefont {White},
  \citenamefont {R{\o}nnow}, \citenamefont {Dewhurst}, \citenamefont
  {Mochizuki}, \citenamefont {Yanai} \emph {et~al.}}]{kezsmarki2015}%
  \BibitemOpen
  \bibfield  {author} {\bibinfo {author} {\bibfnamefont {I.}~\bibnamefont
  {K{\'e}zsm{\'a}rki}}, \bibinfo {author} {\bibfnamefont {S.}~\bibnamefont
  {Bord{\'a}cs}}, \bibinfo {author} {\bibfnamefont {P.}~\bibnamefont {Milde}},
  \bibinfo {author} {\bibfnamefont {E.}~\bibnamefont {Neuber}}, \bibinfo
  {author} {\bibfnamefont {L.}~\bibnamefont {Eng}}, \bibinfo {author}
  {\bibfnamefont {J.}~\bibnamefont {White}}, \bibinfo {author} {\bibfnamefont
  {H.~M.}\ \bibnamefont {R{\o}nnow}}, \bibinfo {author} {\bibfnamefont
  {C.}~\bibnamefont {Dewhurst}}, \bibinfo {author} {\bibfnamefont
  {M.}~\bibnamefont {Mochizuki}}, \bibinfo {author} {\bibfnamefont
  {K.}~\bibnamefont {Yanai}},  \emph {et~al.},\ }\href@noop {} {\bibfield
  {journal} {\bibinfo  {journal} {Nature Materials}\ }\textbf {\bibinfo
  {volume} {14}},\ \bibinfo {pages} {1116} (\bibinfo {year}
  {2015})}\BibitemShut {NoStop}%
\bibitem [{\citenamefont {Tonomura}\ \emph {et~al.}(2012)\citenamefont
  {Tonomura}, \citenamefont {Yu}, \citenamefont {Yanagisawa}, \citenamefont
  {Matsuda}, \citenamefont {Onose}, \citenamefont {Kanazawa}, \citenamefont
  {Park},\ and\ \citenamefont {Tokura}}]{tonomura2012}%
  \BibitemOpen
  \bibfield  {author} {\bibinfo {author} {\bibfnamefont {A.}~\bibnamefont
  {Tonomura}}, \bibinfo {author} {\bibfnamefont {X.}~\bibnamefont {Yu}},
  \bibinfo {author} {\bibfnamefont {K.}~\bibnamefont {Yanagisawa}}, \bibinfo
  {author} {\bibfnamefont {T.}~\bibnamefont {Matsuda}}, \bibinfo {author}
  {\bibfnamefont {Y.}~\bibnamefont {Onose}}, \bibinfo {author} {\bibfnamefont
  {N.}~\bibnamefont {Kanazawa}}, \bibinfo {author} {\bibfnamefont {H.~S.}\
  \bibnamefont {Park}}, \ and\ \bibinfo {author} {\bibfnamefont
  {Y.}~\bibnamefont {Tokura}},\ }\href@noop {} {\bibfield  {journal} {\bibinfo
  {journal} {Nano Letters}\ }\textbf {\bibinfo {volume} {12}},\ \bibinfo
  {pages} {1673} (\bibinfo {year} {2012})}\BibitemShut {NoStop}%
\bibitem [{\citenamefont {Wilson}\ \emph {et~al.}(2014)\citenamefont {Wilson},
  \citenamefont {Butenko}, \citenamefont {Bogdanov},\ and\ \citenamefont
  {Monchesky}}]{wilson2014}%
  \BibitemOpen
  \bibfield  {author} {\bibinfo {author} {\bibfnamefont {M.~N.}\ \bibnamefont
  {Wilson}}, \bibinfo {author} {\bibfnamefont {A.~B.}\ \bibnamefont {Butenko}},
  \bibinfo {author} {\bibfnamefont {A.~N.}\ \bibnamefont {Bogdanov}}, \ and\
  \bibinfo {author} {\bibfnamefont {T.~L.}\ \bibnamefont {Monchesky}},\
  }\href@noop {} {\bibfield  {journal} {\bibinfo  {journal} {Physical Review
  B}\ }\textbf {\bibinfo {volume} {89}},\ \bibinfo {pages} {094411} (\bibinfo
  {year} {2014})}\BibitemShut {NoStop}%
\bibitem [{\citenamefont {Heinze}\ \emph {et~al.}(2011)\citenamefont {Heinze},
  \citenamefont {Von~Bergmann}, \citenamefont {Menzel}, \citenamefont {Brede},
  \citenamefont {Kubetzka}, \citenamefont {Wiesendanger}, \citenamefont
  {Bihlmayer},\ and\ \citenamefont {Bl{\"u}gel}}]{heinze2011}%
  \BibitemOpen
  \bibfield  {author} {\bibinfo {author} {\bibfnamefont {S.}~\bibnamefont
  {Heinze}}, \bibinfo {author} {\bibfnamefont {K.}~\bibnamefont
  {Von~Bergmann}}, \bibinfo {author} {\bibfnamefont {M.}~\bibnamefont
  {Menzel}}, \bibinfo {author} {\bibfnamefont {J.}~\bibnamefont {Brede}},
  \bibinfo {author} {\bibfnamefont {A.}~\bibnamefont {Kubetzka}}, \bibinfo
  {author} {\bibfnamefont {R.}~\bibnamefont {Wiesendanger}}, \bibinfo {author}
  {\bibfnamefont {G.}~\bibnamefont {Bihlmayer}}, \ and\ \bibinfo {author}
  {\bibfnamefont {S.}~\bibnamefont {Bl{\"u}gel}},\ }\href@noop {} {\bibfield
  {journal} {\bibinfo  {journal} {Nature Physics}\ }\textbf {\bibinfo {volume}
  {7}},\ \bibinfo {pages} {713} (\bibinfo {year} {2011})}\BibitemShut {NoStop}%
\bibitem [{\citenamefont {Yu}\ \emph {et~al.}(2012)\citenamefont {Yu},
  \citenamefont {Kanazawa}, \citenamefont {Zhang}, \citenamefont {Nagai},
  \citenamefont {Hara}, \citenamefont {Kimoto}, \citenamefont {Matsui},
  \citenamefont {Onose},\ and\ \citenamefont {Tokura}}]{yu2012}%
  \BibitemOpen
  \bibfield  {author} {\bibinfo {author} {\bibfnamefont {X.~Z.}\ \bibnamefont
  {Yu}}, \bibinfo {author} {\bibfnamefont {N.}~\bibnamefont {Kanazawa}},
  \bibinfo {author} {\bibfnamefont {W.~Z.}\ \bibnamefont {Zhang}}, \bibinfo
  {author} {\bibfnamefont {T.}~\bibnamefont {Nagai}}, \bibinfo {author}
  {\bibfnamefont {T.}~\bibnamefont {Hara}}, \bibinfo {author} {\bibfnamefont
  {K.}~\bibnamefont {Kimoto}}, \bibinfo {author} {\bibfnamefont
  {Y.}~\bibnamefont {Matsui}}, \bibinfo {author} {\bibfnamefont
  {Y.}~\bibnamefont {Onose}}, \ and\ \bibinfo {author} {\bibfnamefont
  {Y.}~\bibnamefont {Tokura}},\ }\href@noop {} {\bibfield  {journal} {\bibinfo
  {journal} {Nature Communications}\ }\textbf {\bibinfo {volume} {3}},\
  \bibinfo {pages} {988} (\bibinfo {year} {2012})}\BibitemShut {NoStop}%
\bibitem [{\citenamefont {White}\ \emph {et~al.}(2014)\citenamefont {White},
  \citenamefont {Pr{\v{s}}a}, \citenamefont {Huang}, \citenamefont {Omrani},
  \citenamefont {{\v{Z}}ivkovi{\'c}}, \citenamefont {Bartkowiak}, \citenamefont
  {Berger}, \citenamefont {Magrez}, \citenamefont {Gavilano}, \citenamefont
  {Nagy} \emph {et~al.}}]{white2014}%
  \BibitemOpen
  \bibfield  {author} {\bibinfo {author} {\bibfnamefont {J.}~\bibnamefont
  {White}}, \bibinfo {author} {\bibfnamefont {K.}~\bibnamefont {Pr{\v{s}}a}},
  \bibinfo {author} {\bibfnamefont {P.}~\bibnamefont {Huang}}, \bibinfo
  {author} {\bibfnamefont {A.~A.}\ \bibnamefont {Omrani}}, \bibinfo {author}
  {\bibfnamefont {I.}~\bibnamefont {{\v{Z}}ivkovi{\'c}}}, \bibinfo {author}
  {\bibfnamefont {M.}~\bibnamefont {Bartkowiak}}, \bibinfo {author}
  {\bibfnamefont {H.}~\bibnamefont {Berger}}, \bibinfo {author} {\bibfnamefont
  {A.}~\bibnamefont {Magrez}}, \bibinfo {author} {\bibfnamefont
  {J.}~\bibnamefont {Gavilano}}, \bibinfo {author} {\bibfnamefont
  {G.}~\bibnamefont {Nagy}},  \emph {et~al.},\ }\href@noop {} {\bibfield
  {journal} {\bibinfo  {journal} {Physical Review Letters}\ }\textbf {\bibinfo
  {volume} {113}},\ \bibinfo {pages} {107203} (\bibinfo {year}
  {2014})}\BibitemShut {NoStop}%
\bibitem [{\citenamefont {Dzyaloshinski}(1958)}]{D}%
  \BibitemOpen
  \bibfield  {author} {\bibinfo {author} {\bibfnamefont {I.~E.}\ \bibnamefont
  {Dzyaloshinski}},\ }\href@noop {} {\bibfield  {journal} {\bibinfo  {journal}
  {J. Phys. Chem. Solids}\ }\textbf {\bibinfo {volume} {4}},\ \bibinfo {pages}
  {241} (\bibinfo {year} {1958})}\BibitemShut {NoStop}%
\bibitem [{\citenamefont {Moriya}(1960)}]{M}%
  \BibitemOpen
  \bibfield  {author} {\bibinfo {author} {\bibfnamefont {T.}~\bibnamefont
  {Moriya}},\ }\href@noop {} {\bibfield  {journal} {\bibinfo  {journal} {Phys.
  Rev}\ }\textbf {\bibinfo {volume} {120}},\ \bibinfo {pages} {91} (\bibinfo
  {year} {1960})}\BibitemShut {NoStop}%
\bibitem [{\citenamefont {Bak}\ and\ \citenamefont {Jensen}(1980)}]{bak1980}%
  \BibitemOpen
  \bibfield  {author} {\bibinfo {author} {\bibfnamefont {P.}~\bibnamefont
  {Bak}}\ and\ \bibinfo {author} {\bibfnamefont {M.~H.}\ \bibnamefont
  {Jensen}},\ }\href@noop {} {\bibfield  {journal} {\bibinfo  {journal}
  {Journal of Physics C: Solid State Physics}\ }\textbf {\bibinfo {volume}
  {13}},\ \bibinfo {pages} {L881} (\bibinfo {year} {1980})}\BibitemShut
  {NoStop}%
\bibitem [{\citenamefont {Buhrandt}\ and\ \citenamefont
  {Fritz}(2013)}]{buhrandt2013}%
  \BibitemOpen
  \bibfield  {author} {\bibinfo {author} {\bibfnamefont {S.}~\bibnamefont
  {Buhrandt}}\ and\ \bibinfo {author} {\bibfnamefont {L.}~\bibnamefont
  {Fritz}},\ }\href@noop {} {\bibfield  {journal} {\bibinfo  {journal}
  {Physical Review B}\ }\textbf {\bibinfo {volume} {88}},\ \bibinfo {pages}
  {195137} (\bibinfo {year} {2013})}\BibitemShut {NoStop}%
\bibitem [{\citenamefont {Bogdanov}\ and\ \citenamefont
  {Yablonskii}(1989)}]{bogdanov1989}%
  \BibitemOpen
  \bibfield  {author} {\bibinfo {author} {\bibfnamefont {A.~N.}\ \bibnamefont
  {Bogdanov}}\ and\ \bibinfo {author} {\bibfnamefont {D.~A.}\ \bibnamefont
  {Yablonskii}},\ }\href@noop {} {\bibfield  {journal} {\bibinfo  {journal}
  {Zh. Eksp. Teor. Fiz}\ }\textbf {\bibinfo {volume} {95}},\ \bibinfo {pages}
  {182} (\bibinfo {year} {1989})}\BibitemShut {NoStop}%
\bibitem [{\citenamefont {Bogdanov}\ and\ \citenamefont
  {Hubert}(1994)}]{bogdanov1994}%
  \BibitemOpen
  \bibfield  {author} {\bibinfo {author} {\bibfnamefont {A.~N.}\ \bibnamefont
  {Bogdanov}}\ and\ \bibinfo {author} {\bibfnamefont {A.}~\bibnamefont
  {Hubert}},\ }\href@noop {} {\bibfield  {journal} {\bibinfo  {journal}
  {Journal of Magnetism and Magnetic Materials}\ }\textbf {\bibinfo {volume}
  {138}},\ \bibinfo {pages} {255} (\bibinfo {year} {1994})}\BibitemShut
  {NoStop}%
\bibitem [{\citenamefont {R{\"o}{\ss}ler}\ \emph {et~al.}(2006)\citenamefont
  {R{\"o}{\ss}ler}, \citenamefont {Bogdanov},\ and\ \citenamefont
  {Pfleiderer}}]{rossler2006}%
  \BibitemOpen
  \bibfield  {author} {\bibinfo {author} {\bibfnamefont {U.~K.}\ \bibnamefont
  {R{\"o}{\ss}ler}}, \bibinfo {author} {\bibfnamefont {A.~N.}\ \bibnamefont
  {Bogdanov}}, \ and\ \bibinfo {author} {\bibfnamefont {C.}~\bibnamefont
  {Pfleiderer}},\ }\href@noop {} {\bibfield  {journal} {\bibinfo  {journal}
  {Nature}\ }\textbf {\bibinfo {volume} {442}},\ \bibinfo {pages} {797}
  (\bibinfo {year} {2006})}\BibitemShut {NoStop}%
\bibitem [{\citenamefont {Butenko}\ \emph {et~al.}(2010)\citenamefont
  {Butenko}, \citenamefont {Leonov}, \citenamefont {R{\"o}{\ss}ler},\ and\
  \citenamefont {Bogdanov}}]{butenko2010}%
  \BibitemOpen
  \bibfield  {author} {\bibinfo {author} {\bibfnamefont {A.~B.}\ \bibnamefont
  {Butenko}}, \bibinfo {author} {\bibfnamefont {A.~A.}\ \bibnamefont {Leonov}},
  \bibinfo {author} {\bibfnamefont {U.~K.}\ \bibnamefont {R{\"o}{\ss}ler}}, \
  and\ \bibinfo {author} {\bibfnamefont {A.~N.}\ \bibnamefont {Bogdanov}},\
  }\href@noop {} {\bibfield  {journal} {\bibinfo  {journal} {Physical Review
  B}\ }\textbf {\bibinfo {volume} {82}},\ \bibinfo {pages} {052403} (\bibinfo
  {year} {2010})}\BibitemShut {NoStop}%
\bibitem [{\citenamefont {R{\"o}{\ss}ler}\ \emph {et~al.}(2011)\citenamefont
  {R{\"o}{\ss}ler}, \citenamefont {Leonov},\ and\ \citenamefont
  {Bogdanov}}]{rossler2011}%
  \BibitemOpen
  \bibfield  {author} {\bibinfo {author} {\bibfnamefont {U.~K.}\ \bibnamefont
  {R{\"o}{\ss}ler}}, \bibinfo {author} {\bibfnamefont {A.~A.}\ \bibnamefont
  {Leonov}}, \ and\ \bibinfo {author} {\bibfnamefont {A.~N.}\ \bibnamefont
  {Bogdanov}},\ }in\ \href@noop {} {\emph {\bibinfo {booktitle} {Journal of
  Physics: Conference Series}}},\ Vol.\ \bibinfo {volume} {303}\ (\bibinfo
  {organization} {IOP Publishing},\ \bibinfo {year} {2011})\ p.\ \bibinfo
  {pages} {012105}\BibitemShut {NoStop}%
\bibitem [{\citenamefont {Adams}\ \emph {et~al.}(2011)\citenamefont {Adams},
  \citenamefont {M{\"u}hlbauer}, \citenamefont {Pfleiderer}, \citenamefont
  {Jonietz}, \citenamefont {Bauer}, \citenamefont {Neubauer}, \citenamefont
  {Georgii}, \citenamefont {B{\"o}ni}, \citenamefont {Keiderling},
  \citenamefont {Everschor} \emph {et~al.}}]{adams2011}%
  \BibitemOpen
  \bibfield  {author} {\bibinfo {author} {\bibfnamefont {T.}~\bibnamefont
  {Adams}}, \bibinfo {author} {\bibfnamefont {S.}~\bibnamefont
  {M{\"u}hlbauer}}, \bibinfo {author} {\bibfnamefont {C.}~\bibnamefont
  {Pfleiderer}}, \bibinfo {author} {\bibfnamefont {F.}~\bibnamefont {Jonietz}},
  \bibinfo {author} {\bibfnamefont {A.}~\bibnamefont {Bauer}}, \bibinfo
  {author} {\bibfnamefont {A.}~\bibnamefont {Neubauer}}, \bibinfo {author}
  {\bibfnamefont {R.}~\bibnamefont {Georgii}}, \bibinfo {author} {\bibfnamefont
  {P.}~\bibnamefont {B{\"o}ni}}, \bibinfo {author} {\bibfnamefont
  {U.}~\bibnamefont {Keiderling}}, \bibinfo {author} {\bibfnamefont
  {K.}~\bibnamefont {Everschor}},  \emph {et~al.},\ }\href@noop {} {\bibfield
  {journal} {\bibinfo  {journal} {Physical Review Letters}\ }\textbf {\bibinfo
  {volume} {107}},\ \bibinfo {pages} {217206} (\bibinfo {year}
  {2011})}\BibitemShut {NoStop}%
\bibitem [{\citenamefont {Zhang}\ \emph
  {et~al.}(2016{\natexlab{a}})\citenamefont {Zhang}, \citenamefont {Bauer},
  \citenamefont {Burn}, \citenamefont {Milde}, \citenamefont {Neuber},
  \citenamefont {Eng}, \citenamefont {Berger}, \citenamefont {Pfleiderer},
  \citenamefont {van~der Laan},\ and\ \citenamefont {Hesjedal}}]{zhang2016}%
  \BibitemOpen
  \bibfield  {author} {\bibinfo {author} {\bibfnamefont {S.}~\bibnamefont
  {Zhang}}, \bibinfo {author} {\bibfnamefont {A.}~\bibnamefont {Bauer}},
  \bibinfo {author} {\bibfnamefont {D.~M.}\ \bibnamefont {Burn}}, \bibinfo
  {author} {\bibfnamefont {P.}~\bibnamefont {Milde}}, \bibinfo {author}
  {\bibfnamefont {E.}~\bibnamefont {Neuber}}, \bibinfo {author} {\bibfnamefont
  {L.~M.}\ \bibnamefont {Eng}}, \bibinfo {author} {\bibfnamefont
  {H.}~\bibnamefont {Berger}}, \bibinfo {author} {\bibfnamefont
  {C.}~\bibnamefont {Pfleiderer}}, \bibinfo {author} {\bibfnamefont {G.~V.~V.}\
  \bibnamefont {van~der Laan}}, \ and\ \bibinfo {author} {\bibfnamefont
  {T.}~\bibnamefont {Hesjedal}},\ }\href@noop {} {\bibfield  {journal}
  {\bibinfo  {journal} {Nano Letters}\ }\textbf {\bibinfo {volume} {16}},\
  \bibinfo {pages} {3285} (\bibinfo {year} {2016}{\natexlab{a}})}\BibitemShut
  {NoStop}%
\bibitem [{\citenamefont {Zhang}\ \emph
  {et~al.}(2016{\natexlab{b}})\citenamefont {Zhang}, \citenamefont {Bauer},
  \citenamefont {Berger}, \citenamefont {Pfleiderer}, \citenamefont {Van~der
  Laan},\ and\ \citenamefont {Hesjedal}}]{zhang2016prb}%
  \BibitemOpen
  \bibfield  {author} {\bibinfo {author} {\bibfnamefont {S.~L.}\ \bibnamefont
  {Zhang}}, \bibinfo {author} {\bibfnamefont {A.}~\bibnamefont {Bauer}},
  \bibinfo {author} {\bibfnamefont {H.}~\bibnamefont {Berger}}, \bibinfo
  {author} {\bibfnamefont {C.}~\bibnamefont {Pfleiderer}}, \bibinfo {author}
  {\bibfnamefont {G.~V.~V.}\ \bibnamefont {Van~der Laan}}, \ and\ \bibinfo
  {author} {\bibfnamefont {T.}~\bibnamefont {Hesjedal}},\ }\href@noop {}
  {\bibfield  {journal} {\bibinfo  {journal} {Physical Review B}\ }\textbf
  {\bibinfo {volume} {93}},\ \bibinfo {pages} {214420} (\bibinfo {year}
  {2016}{\natexlab{b}})}\BibitemShut {NoStop}%
\bibitem [{\citenamefont {Makino}\ \emph {et~al.}(2017)\citenamefont {Makino},
  \citenamefont {Reim}, \citenamefont {Higashi}, \citenamefont {Okuyama},
  \citenamefont {Sato}, \citenamefont {Nambu}, \citenamefont {Gilbert},
  \citenamefont {Booth}, \citenamefont {Seki},\ and\ \citenamefont
  {Tokura}}]{makino2017}%
  \BibitemOpen
  \bibfield  {author} {\bibinfo {author} {\bibfnamefont {K.}~\bibnamefont
  {Makino}}, \bibinfo {author} {\bibfnamefont {J.~D.}\ \bibnamefont {Reim}},
  \bibinfo {author} {\bibfnamefont {D.}~\bibnamefont {Higashi}}, \bibinfo
  {author} {\bibfnamefont {D.}~\bibnamefont {Okuyama}}, \bibinfo {author}
  {\bibfnamefont {T.~J.}\ \bibnamefont {Sato}}, \bibinfo {author}
  {\bibfnamefont {Y.}~\bibnamefont {Nambu}}, \bibinfo {author} {\bibfnamefont
  {E.~P.}\ \bibnamefont {Gilbert}}, \bibinfo {author} {\bibfnamefont
  {N.}~\bibnamefont {Booth}}, \bibinfo {author} {\bibfnamefont
  {S.}~\bibnamefont {Seki}}, \ and\ \bibinfo {author} {\bibfnamefont
  {Y.}~\bibnamefont {Tokura}},\ }\href@noop {} {\bibfield  {journal} {\bibinfo
  {journal} {Physical Review B}\ }\textbf {\bibinfo {volume} {95}},\ \bibinfo
  {pages} {134412} (\bibinfo {year} {2017})}\BibitemShut {NoStop}%
\bibitem [{\citenamefont {Langner}\ \emph {et~al.}(2014)\citenamefont
  {Langner}, \citenamefont {Roy}, \citenamefont {Mishra}, \citenamefont {Lee},
  \citenamefont {Shi}, \citenamefont {Hossain}, \citenamefont {Chuang},
  \citenamefont {Seki}, \citenamefont {Tokura}, \citenamefont {Kevan} \emph
  {et~al.}}]{langner2014}%
  \BibitemOpen
  \bibfield  {author} {\bibinfo {author} {\bibfnamefont {M.~C.}\ \bibnamefont
  {Langner}}, \bibinfo {author} {\bibfnamefont {S.}~\bibnamefont {Roy}},
  \bibinfo {author} {\bibfnamefont {S.~K.}\ \bibnamefont {Mishra}}, \bibinfo
  {author} {\bibfnamefont {J.~C.~T.}\ \bibnamefont {Lee}}, \bibinfo {author}
  {\bibfnamefont {X.~W.}\ \bibnamefont {Shi}}, \bibinfo {author} {\bibfnamefont
  {M.}~\bibnamefont {Hossain}}, \bibinfo {author} {\bibfnamefont {Y.~D.}\
  \bibnamefont {Chuang}}, \bibinfo {author} {\bibfnamefont {S.}~\bibnamefont
  {Seki}}, \bibinfo {author} {\bibfnamefont {Y.}~\bibnamefont {Tokura}},
  \bibinfo {author} {\bibfnamefont {S.~D.}\ \bibnamefont {Kevan}},  \emph
  {et~al.},\ }\href@noop {} {\bibfield  {journal} {\bibinfo  {journal}
  {Physical Review Letters}\ }\textbf {\bibinfo {volume} {112}},\ \bibinfo
  {pages} {167202} (\bibinfo {year} {2014})}\BibitemShut {NoStop}%
\bibitem [{\citenamefont {Mochizuki}\ \emph {et~al.}(2014)\citenamefont
  {Mochizuki}, \citenamefont {Yu}, \citenamefont {Seki}, \citenamefont
  {Kanazawa}, \citenamefont {Koshibae}, \citenamefont {Zang}, \citenamefont
  {Mostovoy}, \citenamefont {Tokura},\ and\ \citenamefont
  {Nagaosa}}]{mochizuki2014}%
  \BibitemOpen
  \bibfield  {author} {\bibinfo {author} {\bibfnamefont {M.}~\bibnamefont
  {Mochizuki}}, \bibinfo {author} {\bibfnamefont {X.~Z.}\ \bibnamefont {Yu}},
  \bibinfo {author} {\bibfnamefont {S.}~\bibnamefont {Seki}}, \bibinfo {author}
  {\bibfnamefont {N.}~\bibnamefont {Kanazawa}}, \bibinfo {author}
  {\bibfnamefont {W.}~\bibnamefont {Koshibae}}, \bibinfo {author}
  {\bibfnamefont {J.}~\bibnamefont {Zang}}, \bibinfo {author} {\bibfnamefont
  {M.}~\bibnamefont {Mostovoy}}, \bibinfo {author} {\bibfnamefont
  {Y.}~\bibnamefont {Tokura}}, \ and\ \bibinfo {author} {\bibfnamefont
  {N.}~\bibnamefont {Nagaosa}},\ }\href@noop {} {\bibfield  {journal} {\bibinfo
   {journal} {Nature Materials}\ }\textbf {\bibinfo {volume} {13}},\ \bibinfo
  {pages} {241} (\bibinfo {year} {2014})}\BibitemShut {NoStop}%
\bibitem [{\citenamefont {P{\"o}llath}\ \emph {et~al.}(2017)\citenamefont
  {P{\"o}llath}, \citenamefont {Wild}, \citenamefont {Heinen}, \citenamefont
  {Meier}, \citenamefont {Kronseder}, \citenamefont {Tutsch}, \citenamefont
  {Bauer}, \citenamefont {Berger}, \citenamefont {Pfleiderer}, \citenamefont
  {Zweck} \emph {et~al.}}]{pollath2017}%
  \BibitemOpen
  \bibfield  {author} {\bibinfo {author} {\bibfnamefont {S.}~\bibnamefont
  {P{\"o}llath}}, \bibinfo {author} {\bibfnamefont {J.}~\bibnamefont {Wild}},
  \bibinfo {author} {\bibfnamefont {L.}~\bibnamefont {Heinen}}, \bibinfo
  {author} {\bibfnamefont {T.~N.}\ \bibnamefont {Meier}}, \bibinfo {author}
  {\bibfnamefont {M.}~\bibnamefont {Kronseder}}, \bibinfo {author}
  {\bibfnamefont {L.}~\bibnamefont {Tutsch}}, \bibinfo {author} {\bibfnamefont
  {A.}~\bibnamefont {Bauer}}, \bibinfo {author} {\bibfnamefont
  {H.}~\bibnamefont {Berger}}, \bibinfo {author} {\bibfnamefont
  {C.}~\bibnamefont {Pfleiderer}}, \bibinfo {author} {\bibfnamefont
  {J.}~\bibnamefont {Zweck}},  \emph {et~al.},\ }\href@noop {} {\bibfield
  {journal} {\bibinfo  {journal} {Physical Review Letters}\ }\textbf {\bibinfo
  {volume} {118}},\ \bibinfo {pages} {207205} (\bibinfo {year}
  {2017})}\BibitemShut {NoStop}%
\bibitem [{\citenamefont {Bannenberg}\ \emph
  {et~al.}(2016{\natexlab{a}})\citenamefont {Bannenberg}, \citenamefont
  {Kakurai}, \citenamefont {Qian}, \citenamefont {Leli{\`e}vre-Berna},
  \citenamefont {Dewhurst}, \citenamefont {Onose}, \citenamefont {Endoh},
  \citenamefont {Tokura},\ and\ \citenamefont {Pappas}}]{bannenberg2016}%
  \BibitemOpen
  \bibfield  {author} {\bibinfo {author} {\bibfnamefont {L.~J.}\ \bibnamefont
  {Bannenberg}}, \bibinfo {author} {\bibfnamefont {K.}~\bibnamefont {Kakurai}},
  \bibinfo {author} {\bibfnamefont {F.}~\bibnamefont {Qian}}, \bibinfo {author}
  {\bibfnamefont {E.}~\bibnamefont {Leli{\`e}vre-Berna}}, \bibinfo {author}
  {\bibfnamefont {C.~D.}\ \bibnamefont {Dewhurst}}, \bibinfo {author}
  {\bibfnamefont {Y.}~\bibnamefont {Onose}}, \bibinfo {author} {\bibfnamefont
  {Y.}~\bibnamefont {Endoh}}, \bibinfo {author} {\bibfnamefont
  {Y.}~\bibnamefont {Tokura}}, \ and\ \bibinfo {author} {\bibfnamefont
  {C.}~\bibnamefont {Pappas}},\ }\href@noop {} {\bibfield  {journal} {\bibinfo
  {journal} {Physical Review B}\ }\textbf {\bibinfo {volume} {94}},\ \bibinfo
  {pages} {104406} (\bibinfo {year} {2016}{\natexlab{a}})}\BibitemShut
  {NoStop}%
\bibitem [{\citenamefont {Reim}\ \emph {et~al.}(2017)\citenamefont {Reim},
  \citenamefont {Makino}, \citenamefont {Higashi}, \citenamefont {Nambu},
  \citenamefont {Okuyama}, \citenamefont {Sato}, \citenamefont {Gilbert},
  \citenamefont {Booth},\ and\ \citenamefont {Seki}}]{reim2017}%
  \BibitemOpen
  \bibfield  {author} {\bibinfo {author} {\bibfnamefont {J.~D.}\ \bibnamefont
  {Reim}}, \bibinfo {author} {\bibfnamefont {K.}~\bibnamefont {Makino}},
  \bibinfo {author} {\bibfnamefont {D.}~\bibnamefont {Higashi}}, \bibinfo
  {author} {\bibfnamefont {Y.}~\bibnamefont {Nambu}}, \bibinfo {author}
  {\bibfnamefont {D.}~\bibnamefont {Okuyama}}, \bibinfo {author} {\bibfnamefont
  {T.~J.}\ \bibnamefont {Sato}}, \bibinfo {author} {\bibfnamefont {E.~P.}\
  \bibnamefont {Gilbert}}, \bibinfo {author} {\bibfnamefont {N.}~\bibnamefont
  {Booth}}, \ and\ \bibinfo {author} {\bibfnamefont {S.}~\bibnamefont {Seki}},\
  }in\ \href@noop {} {\emph {\bibinfo {booktitle} {Journal of Physics:
  Conference Series}}},\ Vol.\ \bibinfo {volume} {828}\ (\bibinfo
  {organization} {IOP Publishing},\ \bibinfo {year} {2017})\ p.\ \bibinfo
  {pages} {012004}\BibitemShut {NoStop}%
\bibitem [{\citenamefont {Pappas}\ \emph {et~al.}(2017)\citenamefont {Pappas},
  \citenamefont {Bannenberg}, \citenamefont {Leli{\`e}vre-Berna}, \citenamefont
  {Qian}, \citenamefont {Dewhurst}, \citenamefont {Dalgliesh}, \citenamefont
  {Schlagel}, \citenamefont {Lograsso},\ and\ \citenamefont
  {Falus}}]{pappas2017}%
  \BibitemOpen
  \bibfield  {author} {\bibinfo {author} {\bibfnamefont {C.}~\bibnamefont
  {Pappas}}, \bibinfo {author} {\bibfnamefont {L.~J.}\ \bibnamefont
  {Bannenberg}}, \bibinfo {author} {\bibfnamefont {E.}~\bibnamefont
  {Leli{\`e}vre-Berna}}, \bibinfo {author} {\bibfnamefont {F.}~\bibnamefont
  {Qian}}, \bibinfo {author} {\bibfnamefont {C.~D.}\ \bibnamefont {Dewhurst}},
  \bibinfo {author} {\bibfnamefont {R.~M.}\ \bibnamefont {Dalgliesh}}, \bibinfo
  {author} {\bibfnamefont {D.~L.}\ \bibnamefont {Schlagel}}, \bibinfo {author}
  {\bibfnamefont {T.~A.}\ \bibnamefont {Lograsso}}, \ and\ \bibinfo {author}
  {\bibfnamefont {P.}~\bibnamefont {Falus}},\ }\href@noop {} {\bibfield
  {journal} {\bibinfo  {journal} {Physical Review Letters}\ }\textbf {\bibinfo
  {volume} {119}},\ \bibinfo {pages} {047203} (\bibinfo {year}
  {2017})}\BibitemShut {NoStop}%
\bibitem [{Note1()}]{Note1}%
  \BibitemOpen
  \bibinfo {note} {The field homogeneity of the 3D cryomagnet, defined as `the
  maximum field error over a 10 mm diameter spherical volume', is determined by
  the manufacturer as 0.11 \% of the applied magnetic field.}\BibitemShut
  {Stop}%
\bibitem [{Note2()}]{Note2}%
  \BibitemOpen
  \bibinfo {note} {We mounted a field probe on the window of the cryostat to
  check directly the residual field taking into account its decay with
  distance. This field probe confirmed the absence of a sizeable residual
  field. In addition, we made magnetic field scans from negative to positive
  fields to evaluate the residual field. As the plot of the integrated
  scattered intensity vs magnetic field was symmetric and centered around zero,
  a sizable residual field was absent.}\BibitemShut {Stop}%
\bibitem [{foo()}]{footnote}%
  \BibitemOpen
  \href@noop {} {\bibinfo  {journal} {See Supplemental Material for the
  supplemental movies.}\ }\BibitemShut {NoStop}%
\bibitem [{Note3()}]{Note3}%
  \BibitemOpen
\bibfield  {journal} {  }\bibinfo {note} {The scattering patterns of this
  rotation scan are equivalent to those obtained after zero field cooling the
  sample.}\BibitemShut {Stop}%
\bibitem [{\citenamefont {Bannenberg}\ \emph
  {et~al.}(2016{\natexlab{b}})\citenamefont {Bannenberg}, \citenamefont
  {Lefering}, \citenamefont {Kakurai}, \citenamefont {Onose}, \citenamefont
  {Endoh}, \citenamefont {Tokura},\ and\ \citenamefont
  {Pappas}}]{bannenberg2016squid}%
  \BibitemOpen
  \bibfield  {author} {\bibinfo {author} {\bibfnamefont {L.~J.}\ \bibnamefont
  {Bannenberg}}, \bibinfo {author} {\bibfnamefont {A.~J.~E.}\ \bibnamefont
  {Lefering}}, \bibinfo {author} {\bibfnamefont {K.}~\bibnamefont {Kakurai}},
  \bibinfo {author} {\bibfnamefont {Y.}~\bibnamefont {Onose}}, \bibinfo
  {author} {\bibfnamefont {Y.}~\bibnamefont {Endoh}}, \bibinfo {author}
  {\bibfnamefont {Y.}~\bibnamefont {Tokura}}, \ and\ \bibinfo {author}
  {\bibfnamefont {C.}~\bibnamefont {Pappas}},\ }\href@noop {} {\bibfield
  {journal} {\bibinfo  {journal} {Physical Review B}\ }\textbf {\bibinfo
  {volume} {94}},\ \bibinfo {pages} {134433} (\bibinfo {year}
  {2016}{\natexlab{b}})}\BibitemShut {NoStop}%
\bibitem [{\citenamefont {Bauer}\ \emph {et~al.}(2010)\citenamefont {Bauer},
  \citenamefont {Neubauer}, \citenamefont {Franz}, \citenamefont {M{\"u}nzer},
  \citenamefont {Garst},\ and\ \citenamefont {Pfleiderer}}]{bauer2010}%
  \BibitemOpen
  \bibfield  {author} {\bibinfo {author} {\bibfnamefont {A.}~\bibnamefont
  {Bauer}}, \bibinfo {author} {\bibfnamefont {A.}~\bibnamefont {Neubauer}},
  \bibinfo {author} {\bibfnamefont {C.}~\bibnamefont {Franz}}, \bibinfo
  {author} {\bibfnamefont {W.}~\bibnamefont {M{\"u}nzer}}, \bibinfo {author}
  {\bibfnamefont {M.}~\bibnamefont {Garst}}, \ and\ \bibinfo {author}
  {\bibfnamefont {C.}~\bibnamefont {Pfleiderer}},\ }\href@noop {} {\bibfield
  {journal} {\bibinfo  {journal} {Physical Review B}\ }\textbf {\bibinfo
  {volume} {82}},\ \bibinfo {pages} {064404} (\bibinfo {year}
  {2010})}\BibitemShut {NoStop}%
\bibitem [{\citenamefont {Qian}\ \emph {et~al.}(2016)\citenamefont {Qian},
  \citenamefont {Wilhelm}, \citenamefont {Aqeel}, \citenamefont {Palstra},
  \citenamefont {Lefering}, \citenamefont {Br{\"u}ck},\ and\ \citenamefont
  {Pappas}}]{qian2016}%
  \BibitemOpen
  \bibfield  {author} {\bibinfo {author} {\bibfnamefont {F.}~\bibnamefont
  {Qian}}, \bibinfo {author} {\bibfnamefont {H.}~\bibnamefont {Wilhelm}},
  \bibinfo {author} {\bibfnamefont {A.}~\bibnamefont {Aqeel}}, \bibinfo
  {author} {\bibfnamefont {T.~T.~M.}\ \bibnamefont {Palstra}}, \bibinfo
  {author} {\bibfnamefont {A.~J.~E.}\ \bibnamefont {Lefering}}, \bibinfo
  {author} {\bibfnamefont {E.~H.}\ \bibnamefont {Br{\"u}ck}}, \ and\ \bibinfo
  {author} {\bibfnamefont {C.}~\bibnamefont {Pappas}},\ }\href@noop {}
  {\bibfield  {journal} {\bibinfo  {journal} {Physical Review B}\ }\textbf
  {\bibinfo {volume} {94}},\ \bibinfo {pages} {064418} (\bibinfo {year}
  {2016})}\BibitemShut {NoStop}%
\end{thebibliography}%


\end{document}